\documentclass[12pt,a4paper]{article}
\pdfoutput=1
\usepackage{amsmath}
\usepackage[T1]{fontenc}
\usepackage[utf8]{inputenc}
\usepackage{authblk}

\usepackage{hyperref}
\usepackage{graphicx}
\usepackage{fancybox}
\usepackage{color}
\usepackage{epsfig,graphicx}

\title{   Giant D5 Brane Holographic  Hall State}

 \author[1]{Charlotte Kristjansen}
\author[1,2]{Gordon W. Semenoff}
 
\affil[1]{Niels Bohr Institute, Copenhagen University,
Blegdamsvej 17,   2100 Copenhagen \O, Denmark}
\affil[2]{Department of Physics and Astronomy, University of British Columbia,  Vancouver, BC Canada V6T 1Z1}

\date{} 
\begin{document}

\maketitle

\begin{abstract}
  We find a new  holographic description of strongly coupled defect field theories using probe D5 branes.   We consider 
a system where a large number of probe branes, which are asymptotically D5 branes, blow up into a D7 brane suspended in the bulk 
of anti-de Sitter space.   For a particular ratio of charge density to external magnetic field, 
so that the Landau level filling fraction per color is equal to one, 
the D7 brane exhibits an incompressible charge-gapped state with one unit of integer quantized Hall conductivity. 
The detailed configuration as well as ungapped, compressible configurations for a range of parameters near the gapped one are  found by
solving the D5 and D7 brane embedding equations numerically and the D7 is shown to be  preferred over the D5 by comparing their energies. 
We then find integer quantum Hall states with higher filling fractions as a stack of D5 branes which blow up to multiple D7 branes where each
D7 brane has filling fraction one.  We find indications that the $\nu$ D7 branes describing the filling fraction $\nu$ 
state are coincident with a residual
SU($\nu$) symmetry when $\nu$ is a divisor of the total number of D5 branes. 
We examine the issue of stability of the larger filling fraction Hall states. We argue that, in the D7 brane phase, chiral symmetry restoration could
be a first order phase transition.  
\end{abstract}

\section{Introduction and discussion}

The  D3-D5 brane system  is one of the best studied examples of top-down holography using probe branes \cite{Karch:2000gx}-\cite{Brattan:2012nb}.  A stack of $N_5$ coincident and appropriately oriented D5 branes is 
placed in the vicinity of a stack of $N$ coincident D3 branes. In the limit where the number of D3 branes N is taken large and the string coupling $g_s$ small
so that  $\lambda=4\pi g_sN$ is held constant, the D3 branes are replaced by the $AdS_5\times S^5$ background spacetime
and the D5-branes are a probe embedded in this background.   The probe limit requires that $N_5<<N$.  
The string theory is weakly coupled in the limit where $\lambda$
is large. In this limit, the mathematical problem of understanding the probe branes reduces to understanding the embedding of
their worldvolume in $AdS_5\times S^5$. 

The field theory dual of the large $N$ limit of string theory on $AdS_5\times S^5$ is the planar limit of 
${\mathcal N}=4$ supersymmetric Yang Mills theory with 't Hooft coupling $\lambda$.  The role of the probe branes is to introduce
fundamental representation degrees of freedom.  In the D3-D5 system, these degrees of freedom constitute a supersymmetric hypermultiplet which occupies a 2+1-dimensional defect embedded in 3+1-dimensional spacetime. This system has a solution with superconformal symmetry which preserves half of the supersymmetries of the ${\mathcal N}=4$ theory and the conformal symmetry of the defect. The probe brane worldvolume is the space $AdS_4\times S^2$ which has bosonic symmetries $SO(3,2)\times SO(3)\times SO(3)$. When there are $N_5$ coincident D5 branes, there is  a $U(N_5)$ flavor symmetry
in the field theory dual. 

When temperature, an external magnetic field and charge density are introduced, the D3-D5 system has a rich phase diagram \cite{Evans:2010hi}.  
For example, at zero temperature and density, the presence of a constant external magnetic field results in  spontaneous breaking of an SO(3) chiral symmetry  of the dual field theory.  This is an example of magnetic catalysis of chiral symmetry breaking, where the tendency of interactions to 
break chiral symmetry is greatly enhanced in the presence of an external magnetic field \cite{cat0}-\cite{Filev:2012ch}.  
Due to chiral symmetry breaking, the field theory spectrum gets a charge gap.  In the holographic theory, this is seen as the probe brane taking up a "Minkowski embedding'', where it still approaches $AdS_4$ near the boundary of $AdS_5$, but it is modified in the 
bulk and it caps off before it reaches the Poincare horizon.  The charge gap is
seen in the fact that the D3-D5 strings, whose low energy modes are the U(1) charged fundamental representation matter degrees of freedom, must be suspended between the D5 worldvolume and the Poincare horizon.  When the worldvolume does not reach the horizon, the D3-D5 strings must
have a minimum length and therefore they have an energy gap.  

When a charge density is introduced in addition to the magnetic field, there is a low density phase which still exhibits dynamical chiral symmetry breaking. 
In all cases, when the charge density is nonzero, the D5 brane embedding must be a ``black hole embedding'', rather than Minkowski embedding.\footnote{The term ``black hole embedding'' which normally used to describe  a probe brane with worldvolume which reaches an 
$AdS_5$ black hole horizon as well as intersecting the boundary of $AdS_5$,  Here we shall use this term, 
even though we shall deal solely with the zero temperature 
limit where there is no black hole.  We use it to refer to probe brane whose worldvolume reaches  the Poincare horizon of the Poincare coordinate patch of $AdS_5$.} 
It must stretch from the boundary of $AdS_5$ to the Poincare horizon.
This is due to the fact that it carries worldvolume electric flux.  It is not possible for the worldvolume to pinch off smoothly
unless there is a sink to absorb the electric flux.
Fundamental strings suspended between the worldvolume and the horizon could provide such a sink.  However,
fundamental strings with the necessary configuration always have a larger tension than the D5-brane, 
and they would pull the D5 brane to the horizon.  Indeed, the numerical solutions
for the worldvolume geometry do exhibit a spike which resembles such a fundamental string funnel. 
A black hole embedding is thus a compressible state.  It does not have a charge gap.   
It is somewhat puzzling that the D5 brane states do not have a charge gap and are compressible 
for any non-zero value of the charge density.
In this paper, we examine the question whether the U(1) charged D5 brane can have an incompressible state at all. 
We shall find that, for an interesting range of U(1) charge, the answer is yes and we will construct the corresponding
solutions. 

In rough outline, our solution of this problem is as follows.   We consider $N_5$ D5 branes which enter $AdS_5$ at its boundary. At the boundary the D5 brane worldvolume wraps a 2-sphere $S^2$ embedded in $S^5$ and has an asymptotically $AdS_4$ component embedded
in $AdS_5$.   It also has worldvolume gauge fields which are needed so that the dual field theory has a constant external magnetic field and a constant $U(1)\subset U(N_5)$ charge density.  For simplicity, we will consider only the case where the temperature is zero, although we anticipate that
most of our results could easily be generalized to finite temperature.  
As the AdS radius decreases, the D5 branes blow up via the Myers effect \cite{Myers:1999ps} 
to a D7 brane whose worldvolume wraps a second two 2-sphere $\tilde S^2$ in $S^5$. The second 2-sphere  
carries a Dirac monopole bundle of the D7 worldvolume gauge field with $N_5$ units of monopole charge.  The same coupling of the D7 brane to the Ramond-Ramond 4-form fields which implements the Myers effect also generates a topological  term $\sim F\wedge F\wedge C^{(4)}$ 
in the D7 brane worldvolume gauge theory. This term   allows the
magnetic flux to carry electric charge.  In this way, the electric charge dissolves into the D7 brane.  For a certain value of
the charge density, all of the charge can be dissolved and the D7 brane is allowed to cap off at a finite AdS radius, that is, to have a Minkowski embedding.  The result is an incompressible state with a charge gap. The filling fraction  (the dimensionless ratio of charge density to magnetic field) at which this occurs is precisely that of a $\nu=1$ integer quantum Hall state.  If the filling fraction $\nu$ is close to, but not equal to one, the D7 brane still exists, but it has a different character as it has residual charge and it must therefore  have a black hole
embedding which reaches the Poincare horizon.  We are able to confirm this by numerical solution of the 
embedding equations for the D7 brane where we find the Minkowski embedding at $\nu=1$ and black hole embeddings for the
D5 brane with any nonzero charge density and the D7 brane when $\nu\neq 1$.
We also find that these D7 brane solutions have lower energy than D5 brane solutions with the same parameters, and are therefore favored, at least in an interesting parameter region including $\nu=1$ and perhaps a much larger region which is yet to be explored.   Our analysis is 
most reliable when $N_5
\sim\sqrt{\lambda}$ which is large. Of course, our holographic computations are reliable when $\sqrt{\lambda}$ is large.

Then, we also find higher integer quantized Hall systems by considering, for $\nu=2$, for example, $N_5$ D5 branes where
a subset of them, say 
$N_{5a}$ D5 branes, blow up into one D7 brane and the remaining $N_{5b}=N_{5}-N_{5a}$ 
D5 branes blow up into a second D7 brane and where both of
the latter branes are in the $\nu=1$ charge-gapped incompressible state.  For $\nu=2$, we show that this state is preferred over
both the un-gapped $\nu=2$ D7 brane solution and the D5 brane with the same parameters.  We also find that the symmetric
solution where $N_{5a}=N_{5b}=N_5/2$ seems to be preferred.  This state has a residual $SU(2)$ symmetry, and to have this symmetry
in the strictest sense, $N_5$ must be an even number.  We speculate that there can be 
a large number of stable integer quantum Hall states up to the maximum filling fraction which can be achieved in this way, $\nu=N_5$
although many are outside of the domain where our analysis using probe branes is reliable.  In order for a higher integer filling fraction
to be $SU(\nu)$ symmetric, $\nu$ must be a divisor of $N_5$. 

Our analysis of the incompressible state will be entirely from the D7-brane point of view.  There should be an alternative approach,
which we have not explored yet, where the object is a non-abelian configuration of D5-branes, similar to
the bion which occurs on flat space \cite{malda}\cite{Myers:1999ps}.   
The transverse coordinates of the D5-brane become non-commutative and form a fuzzy sphere.  
This fuzzy sphere is the D5-picture of the additional sphere which is wrapped by the D7 brane worldvolume in our analysis.  The U(1) flux
on this sphere records the number of D5-branes.   
From the D7 brane point of view, the bion spike is just the portion of the D7 brane which
reaches the boundary of $AdS_5$ where the 2-sphere collapses and it resembles a D5 brane.  The source for this spike in
the D5 worldvolume is $N_5$ units of D5 brane
charge which are attached to the D7-brane at its asymptotic boundary.  
One would expect that, like the
case of the bion, the D5 brane analysis is more accurate in the core of the object, that is near the spike,  
and the D7 analysis should be good away from the spike, in the interior of $AdS_5$. It would be interesting
if this could be exploited to better understand some dynamical issues such as how the difference of energies of the D5 and D7 occurs
from that point of view. 
 
\subsection{Weak coupling}

\begin{figure}
\begin{center}
\includegraphics[scale=0.2]{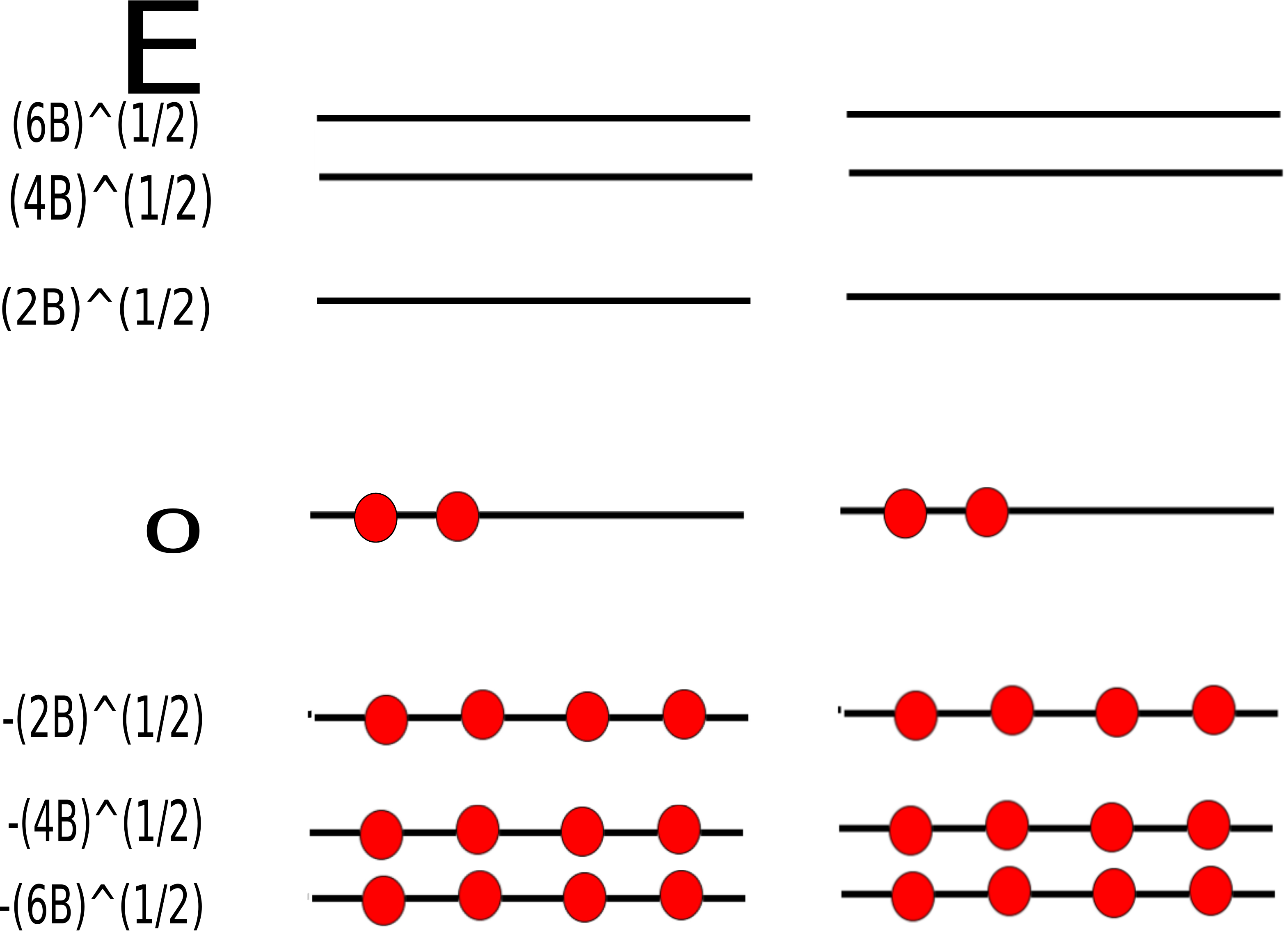}
\end{center}
\caption{\label{fig1} {\footnotesize The Landau level spectrum of an SO(3) doublet of 
massless non-interacting  fermions
in a magnetic field is depicted. The energy levels are the horizontal lines and the vertical axis is the 
single-particle energy.  Only fermionic levels are shown here. We have assumed that $N_5=1$ and we
have suppresed the color degeneracy. The two columns represent the two isospin
states. 
Fermions are depicted by the dots. In the neutral ground state, 
negative energy levels are filled and positive energy levels are empty. The zero energy states
must be half-filled.  In the absence of interactions, all possible fillings of the zero energy states have the same energy.  
This gives rise to  a large degeneracy of the many-body ground state.}}
\end{figure}

It is interesting to examine what one would expect in the weak coupling limit.   Our purpose here is to understand the 
origin of Hall states and the role of the scalar field
which is necessarily present in this supersymmetric theory.    
 If the coupling constant is set to zero, the action of the defect theory is for a $N_5$ D5 branes is known to be \cite{DeWolfe:2001pq}
$$
S_{\rm D}=\int d^3x\left\{|D_\mu q|^2-i\bar\Psi\gamma^\mu D_\mu\Psi\right\},
$$
where $q$ is a complex scalar field which is a spinor representation of $SO(3)$, the isometry group of the worldvolume $S^2$,
and $\Psi$ is a complex
fermion which is in a spinor representation of $\tilde {SO}(3)$, the isometry group of $\tilde S^2$ 
as well as  a spinor of the $SO(2,1)$ Lorentz group. The 2-sphere $S^2$ is the one which was originally wrapped by the D5 brane
and which the D7 brane worldvolume also wraps.  The 2-sphere $\tilde S^2$ is the one which is formed as a fuzzy sphere of D5 branes
and which is wrapped by the D7 worldvolume and contains $N_5$ units of D7 worldvolume magnetic flux. Both $q$ and $\Psi$
transform in the fundamental representation of the bulk SU(N) gauge group and a flavor SU($N_5$) symmetry. They have a color
index which runs from 1 to N and flavor index which runs from 1 to $N_5$. 

It is easy to see that, in the presence of a magnetic field,  this system will break the $SO(3)$ symmetry when there is even 
an infinitesimally weak interaction.  
  In an external $U(1)$ magnetic field, the single particle energy levels of the fermions are the Landau levels 
$E_n=\sqrt{ 2Bn}$ and the scalar field are $\omega_n=\sqrt{(2n+1)B}$ with $n=0,1,2,...$ in both cases.  The levels are infinitely
degenerate.  The density, i.e.~the number of states per unit volume, of a single level is $\frac{B}{2\pi}$.  The degeneracy of each
fermionic Landau level is then $ 2N_5N\frac{B}{2\pi}$ where the factor of 2 is due to the fact that the fermion is an $\tilde {SO(3)}$ spinor, $N_5$
and $N$ come from the numbers of flavors and colors, respectively.  

The bosons have a gap and are not excited at energies less than $\omega_0=\sqrt{B}$.  The fermions, 
on the other hand, have a zero energy mode, $E_0$ which will dominate the low energy properties of the system.  
Moreover, in the neutral ground state of the system, depicted for the case of one flavor, 
$N_5=1$  (and two $\tilde{SO(3)}$ isospin states), in figure 1, 
all negative energy fermion states
must be occupied, positive energy states must be empty and the zero modes should be half-filled.   The latter requirement is due to a charge
conjugation (or particle-hole) symmetry of the spectrum of the Dirac Hamiltonian in a magnetic field.

\begin{figure}
\begin{center}
\includegraphics[scale=0.2]{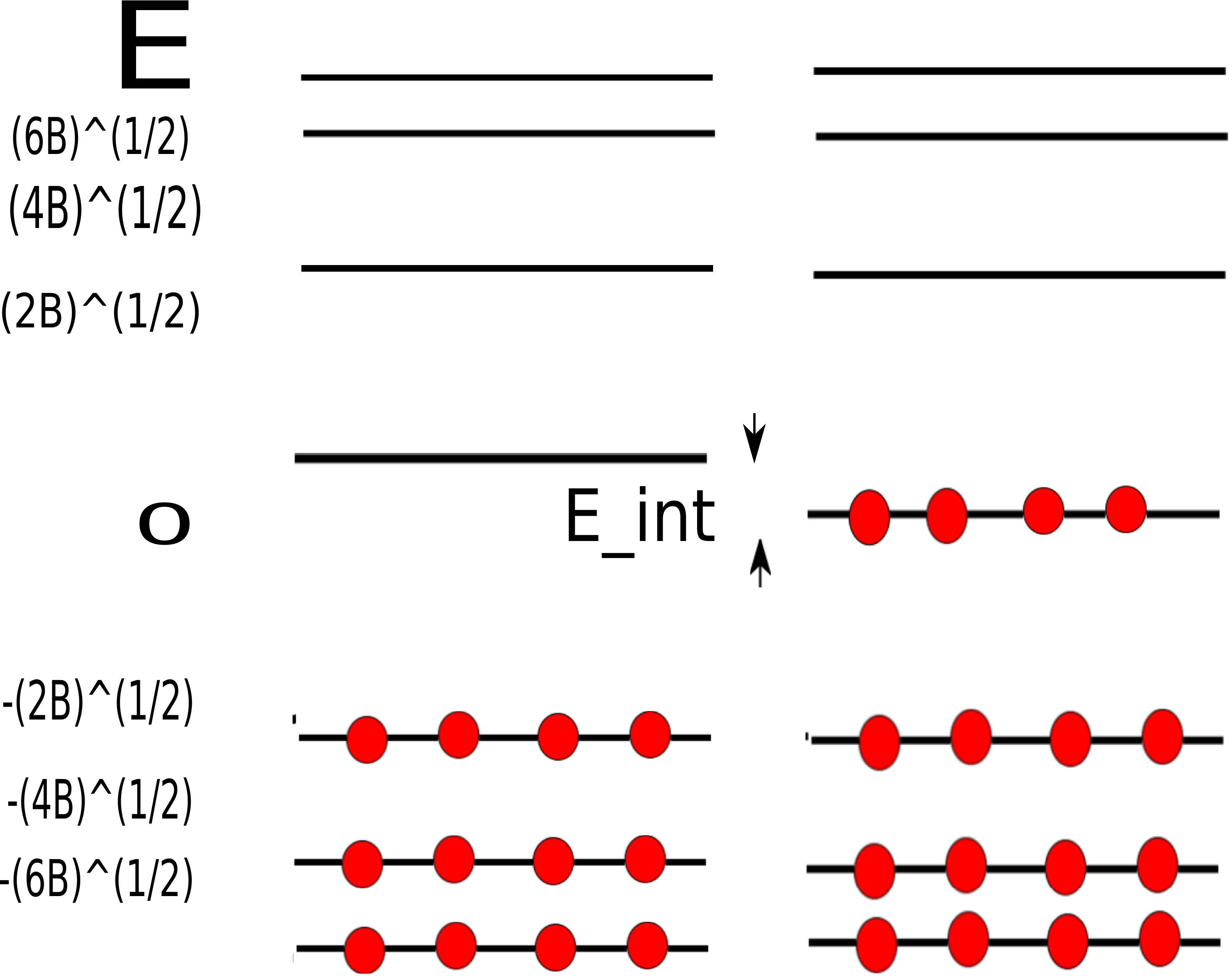}
\end{center}
\caption{\label{fig2} {\footnotesize Magnetic catalysis: with an arbitrarily weak repulsive
interaction (such as the Coulomb interaction), the preferred state is one where all of the zero modes
with one label are occupied to obtain the $\nu=0$ state.   This is the analog of the Hund rule for electrons filling atomic orbitals.  It
can be shown to minimize the exchange energy.  This is a gapped state which, since the density vanishes, has
vanishing Hall conductivity.}}
\end{figure}

An arbitrarily weak interaction will have significant 
influence on the zero modes. The perturbative shift of the energy of states such as in figure \ref{fig1} due to 
a weak repulsive interaction like the Coulomb interaction is known to be minimal for the states depicted in
figure \ref{fig2}, where all of the zero mode fermions take up one of the $\tilde{SO(3)}$ spinor indices \cite{Semenoff:2011ya}.  This is similar to, and
has the same source as the Hund rule for populating electron energy levels of atoms.  In rough terms, 
an interaction like the Coulomb interaction, which is repulsive at all scales, would be minimized by keeping
the fermions as far apart as possible.  Since their wavefunction is totally antisymmetric, this is achieved by making
it as antisymmetric in particle positions as possible, and therefore as symmetric as possible in the other indices.
Symmetry in the flavor indices is maximal if all states have the same index.  This is the symmetry breaking
state depicted in figure \ref{fig2}.
This phenomenon is known in the condensed matter physics literature as quantum Hall 
ferromagnetism and it is the explanation of an experimentally
observed  feature of some two dimensional electron gases \cite{quantumhallferromagnet0}-\cite{quantumhallferromagnet3}.

\begin{figure}
\begin{center}
\includegraphics[scale=0.2]{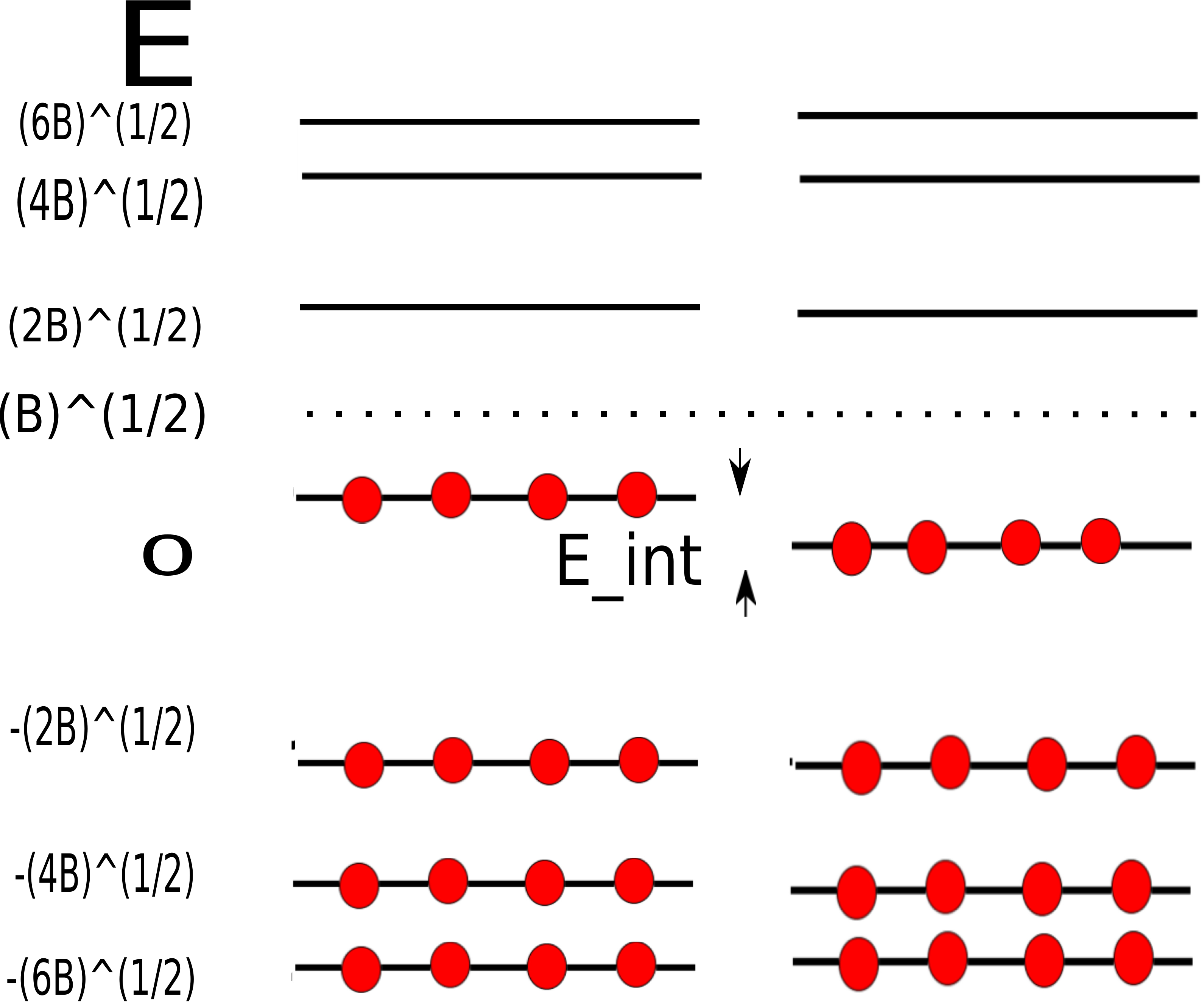}
\end{center}
\caption{\label{fig3} {\footnotesize Beginning with the state depicted in figure \ref{fig2} and increasing the chemical potential 
and charge density eventually results in the $\nu=1$ state that is depicted here.  The energy needed to excite the first bosonic
states is depicted by the dashed line.  Further increasing the charge will simply increase the number of Bosons excited.  
Thus, the $\nu=1$ state is the only quantum Hall state in the spectrum.}}
\end{figure}

Now, assume that interactions have generated a small gap.  First, we observe that this breaks the $\tilde{SO(3)}$ chiral symmetry and
it makes the neutral state a gapped
state, where otherwise, if the symmetry was not broken, it would have been ungapped.  What will the
system look like if we now vary the density?  If we begin with the neutral state of figure \ref{fig2} and add fermions we immediately have
an ungapped state with the second set of would-be zero modes partially filled.  Then, if we continue to add fermions, we will eventually 
fill all of the remaining would-be zero
modes which are now at a small positive energy.  Then the completely filled state, depicted in figure \ref{fig3}, also has a gap.  
This state is an integer quantum Hall state\footnote{Note that, in this translational invariant quantum field theory, even though we have gapped states for
certain discrete values of the charge density, the usual signature of the quantized Hall effect, the Hall plateaus in the graph of Hall conductivity versus filling fraction will not occur.  
Of course, localized states, which are absent here, would be needed in
order for integer quantum Hall plateaus to form and for this system to have a conventional quantum Hall effect. Here, we nevertheless
refer to the charge gapped state 
obtained by tuning the charge density to the appropriate value as a ``quantum Hall state''. It is distinguished from other states by virtue of being
incompressible and by the existence of 
a charge gap. } with filling
fraction $\nu=1$, where
\footnote{Note that this filling fraction is defined with a factor of the number of colors $N$ in the denominator.   The first intger filling fraction $\nu=1$
refers to a Landau level which is completely filled with baryons, where the baryon is a composite of $N$ fundamental representation quarks and has U(1) charge N.  
This color degeneracy is unavoidable here and it is natural if the theory is confining, 
as one would expect when the chiral symmetry is spontaneously broken, as it is here. } 
\begin{align}\label{definenu}
\nu=\frac{2\pi}{N}\frac{\rho}{B}.
\end{align}
  The color degeneracy ($SU(N)$ of $ {\cal N}=4$ Yang-Mills theory) is explicit in our definition of filling fraction.

If we continue to increase the density, the next available energy states are where we excite bosons whose gap is $\sqrt{B}$. Then there can be no
further Hall states as all further charge can be accomodated by Bosons which do not have a Pauli principle.  
If we decreased, rather than increased the charge density, particle-hole symmetry tells us that we would also find an incompressible state at $\nu=-1$. 
Thus,  at weak coupling, when there is only one flavor of fermions, and because of the presence of charged scalar fields, 
there are only three incompressible states with $\nu=1,0,-1$ and only two Hall states with $\nu=1,-1$.  
This changes if there are multiple flavors. 

\begin{figure}
\begin{center}
\includegraphics[scale=0.2]{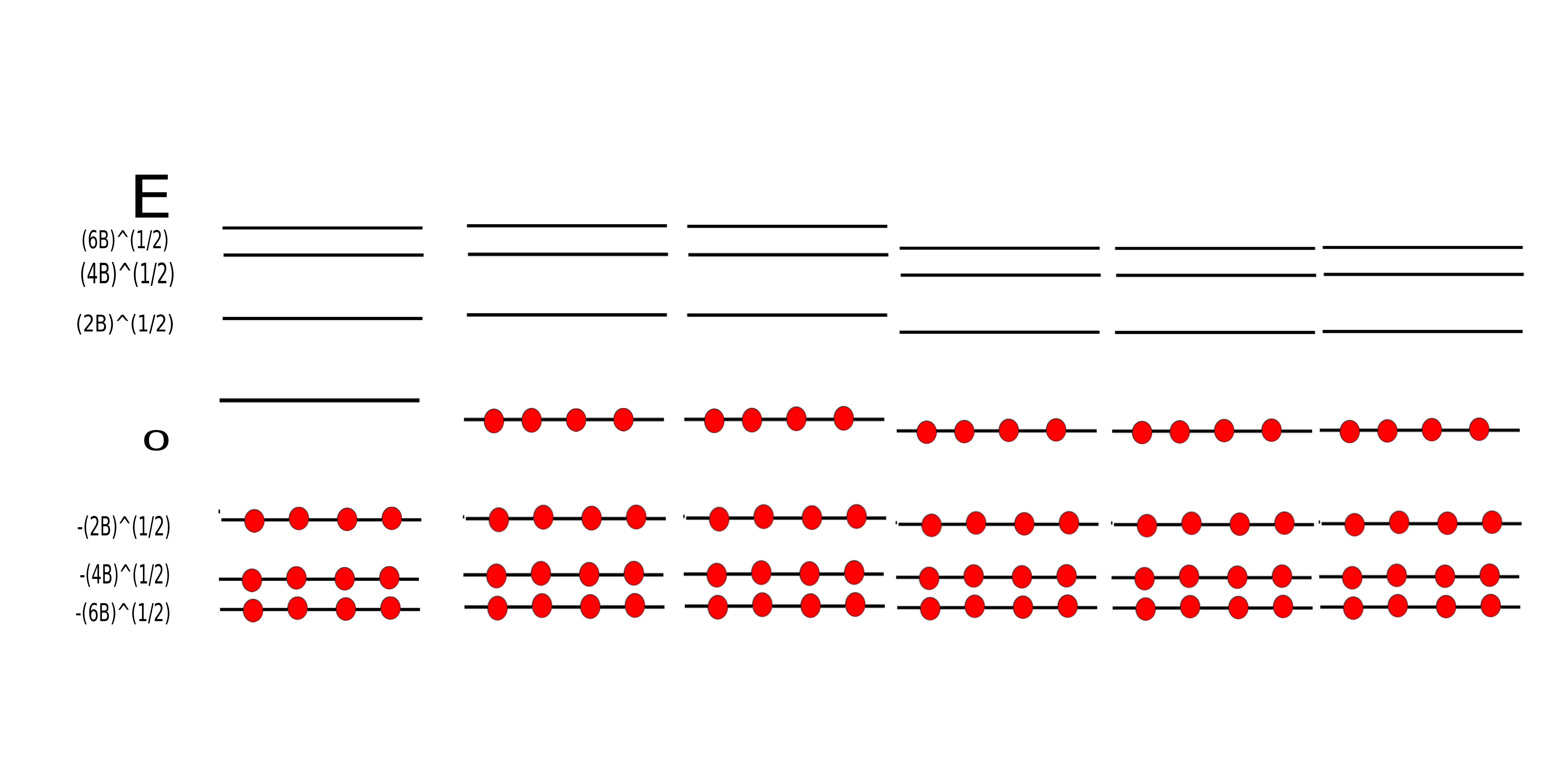}
\end{center}
\caption{\label{fig4} {\footnotesize For $N_5=3$, the possible Hall states have
filling fractions $\nu=1,2,3$. The state with
$\nu=2$ is depicted.   There is an energy gap between the fermions in this state and those
which must be added to continue filling the remaining zero modes.     }}
\end{figure}

For multiple D-branes with $N_5$ flavors of fermions, there are similar arguments to quantum Hall ferromagnetism
for a complete resolution of the degeneracy of 
the zero modes so that filling an $SO(3)$ state for each flavor (with fermins or holes) could potentially yield $2N_5$  
quantum Hall states.  However, the number of these
states which actually become Hall states depends on how many fit under the threshold for creating a scalar particle. At very weak coupling
the number is $ N_5$ levels with filling fractions $\pm\nu= 1,2,..., N_5$ and Hall conductivities $\pm\sigma_{xy}=
\frac{1}{2\pi}, \frac{2}{2\pi},...,\frac{ N_5}{2\pi}$.  An example with $N_5=3$ is depicted in figure \ref{fig4}.

In effective field theory, the Hall conductivity is encoded in a Chern-Simons term.  If, in one of the gapped states, we introduce a background U(1)
gauge field and then we integrate out the matter fields, the action has a Chern Simons term
$$
S_{\rm eff.}=\frac{\sigma_{xy}}{2}\int d^3x ~\epsilon^{\mu\nu\lambda}A_\mu\partial_\nu A_\lambda~+~  \ldots
$$
When there is a charge gap,  the Chern-Simons term does not renormalize beyond one loop 
\cite{Coleman:1985zi}\cite{Semenoff:1988ep}\cite{Lykken:1991xs}. 
It is also known that when there is no gap, it can be corrected at higher loops \cite{Semenoff:1988ep}.  At one loop, the scalar field   does
not contribute and the fermion  gives the result above \cite{Niemi:1983rq}\cite{Redlich:1983kn}. 

 The no-renormalization theorem suggests that the filling fraction  and Hall conductivity of a gapped state 
could be exact and identical to its value in the weak coupling limit.   It is interesting to
check this idea at strong coupling. In this paper, we indeed find, first of all, the Hall state  with the lowest integer Hall
conductivity at filling fraction $\nu=1$, and then others with $\nu=2,3,\ldots$ the total number limited by dynamical issues, but
bounded from above to be less than or equal to $N_5$.  

One might speculate that these states are strong coupling extrapolations of the ones that are 
seen in the weak coupling limit.  It is interesting that the upper limit on the strong coupling side comes from the D brane geometry. 
  We shall also see that the maximum number of Hall states is $N_5$, coming from $N_5$ D5 branes blowing up into
$N_5$ D7 branes.  However, it is very possible that only a much smaller number of these states are stable.  The geometric description of
the blow-up as a classical sphere with magnetic flux, which our analysis relies on, is also not accurate unless the number of D5-branes 
which blow up to a D7 is large.

   \subsection{D7$^\prime$}

The D7 brane configuration that we shall use in this paper is closely related to 
that used in a series of papers \cite{Bergman:2010gm}-\cite{Jokela:2012vn} where one
examines a D3-D7 system where the D7 brane worldvolume wraps $S^2\times\tilde S^2\subset S^5$ and
the D7 brane is stabilized by putting U(1) worldvolume magnetic flux on $\tilde S^2$.  This configuration
is called D7$^\prime$ and it has been used to construct a model of the Hall effect by a mechanism which has
some of the ingredients of 
the one that is used here. Both D7 and D7$^\prime$ probe branes have also been used to model some 
phenomena in planar condensed matter
systems \cite{Rey:2008zz}-\cite{Goykhman:2012vy}.

 An important feature of a Hall state is that it is incompressible, that is, it has an 
energy gap for charged excitations.  For a probe brane to be incompressible, it should have
a Minkowski embedding.  However, to be a Hall system, it must also have electric charge and at
first sight these are incompatible, as we have discussed above, when a probe brane carries the worldvolume
gauge field configurations necessary for its field theory dual to have electric charge, it should have a black hole
embedding.  A way around this occurs in the D7$^\prime$ system exactly as it does in the system that we are interested
in in this paper and it   was already used to construct Hall 
states in reference \cite{Bergman:2010gm}. 
In the presence of external magnetic field, the Wess-Zumino terms in the D7 brane action allow electric charge to be dissolved into
the brane worldvolume.  This could occur for discrete values of the ratio of charge density to magnetic field, the filling fraction.  Then, in 
these special cases, the solution for the D7 brane worldvolume is a Minkowski embedding with a charge gap.  These 
were identified as quantized Hall states of D7$^\prime$.  Due to a different choise of boundary conditions, 
the filling fractions are not integers in that case. 
With the types of embedding chosen for D7$^\prime$, some internal magnetic flux on the D7
brane worldvolume was required for stability of the system.  
The flux was also quantized and that gave a quantization condition for
Hall conductivity (which did not match either the integer or known fractional quantum Hall effects).

The difference between D7$^\prime$ and what we do in the present paper 
is that we choose a different boundary condition at the boundary of $AdS_5$. 
The boundary conditions at the boundary of $AdS_5$ must be chosen so that 
the numerator of the last term in the differential equation which describes the 
behavior of the 5-sphere latitude at which the D7 brane is embedded, equation (\ref{equationforpsi}), 
vanishes there (at large $r$ in (\ref{equationforpsi})).  The numerator of that last term is the partial derivative of the expression
in the denominator whose large $r$ behavior is the `potential' 
$$
V(\psi)=4\sin^4\psi(f^2+4\cos^4\psi).
$$
 We thus need the asymptotic value of $\psi(r)$ to approach 
an extremum of this potential.  Moreover, if we want the fluctuations
about the asymptotic value to have two normalizable modes, it is
necessary that this extremum is a maximum.  Simple calculus tells
us that there are either one or two maxima: if $f^2<\frac{1}{2}$, there
are two maxima at
$
\psi=\frac{1}{2}\arccos \left(\frac{1}{2}\left[\sqrt{1-2f^2}-1\right]\right)
$ and $\psi=\frac{\pi}{2}$. 
On the other hand, if $f^2>\frac{1}{2}$, there is only one maximum
at $\psi=\frac{\pi}{2}$.   In reference \cite{Bergman:2010gm}, they used the first 
maximum.  In the present paper, we will consider
of the second maximum at $\psi=\frac{\pi}{2}$.  This as a place where $\cos^2\psi=0$ and
the D7 brane world-volume degenerates to a six dimensional hypersurface
where one of the  worldvolume spheres, $\tilde S^2$, has shrunk to zero  volume. This was a sphere carrying
magnetic flux. When it shrinks to a point it leaves a singularity in the form
of a point Dirac monopole.  This singularity should
be interpreted as $N_5=\frac{\sqrt{\lambda}}{2\pi}f$ D5 branes ending at the D7 brane.   In fact, we interprete it as
a region where a number $N_5$ of D5 branes takes on a non-Abelian configuration and blows up to form the D7 brane.

 \subsection{Outline}

In section 2 we shall review the problem of determining the geometry of a D5 brane embedded in $AdS_5\times S^5$
when the field theory which is its holographic dual has an external magnetic field and charge density. 
In section 3 we discuss the embedding of the D7 brane with the same external magnetic field and charge density. 
In section 4 we discuss our numerical solutions for the D5 and D7 brane configurations as well as their energies. 
Section 5 contains some conclusions.

\section{D3-D5}

We shall study the D3-D5 system in the probe limit where the number of D5 branes $N_5$ is much
smaller than $N$, the number of D3 branes.   The material of this Section is already well known and can be
found elsewhere in the literature.  We include it here for the convenience of the reader, and to fix our conventions and notation
which will be needed later on when we compare the D5 brane with the D7 brane.  
On flat space, the D3 and D5 branes are oriented
as in Table 1.  They overlap in 2+1 dimensions.  
In principle, with multiple D5 branes, their coordinates are 
matrices and the worldvolume gauge fields have $U(N_5)$ gauge group.  We will begin with an analysis
which assumes that the non-Abelian structure plays no role.  The actions for each of the individual D5 branes
are identical and it suffices to study one of them and to multiply the total action by the number of branes $N_5$.

\begin{align}
\nonumber
\boxed{\begin{array}{rcccccccccccl}
  & & x^0 & x^1 & x^2 & x^3 & x^4 & x^5 & x^6 & x^7 & x^8 & x^9 &\\
& D3 & \times & \times & \times & \times & & &  & & & & \\
& D5 & \times & \times & \times &  & \times  & \times & \times & &  & &  \\
& D7 & \times & \times & \times &  & \times  & \times & \times & \times & \times & &   \\
\end{array}}
\nonumber
\\
{\rm\bf Table~1:~D3,~ D5~ and~ D7~orientation}~~~~~~~~~~~~~~~~~~~~~~~~
\nonumber
\end{align}

  We shall be interested in the limit of the string
theory which coincides with the planar limit of the gauge theory and, after the planar limit is 
taken, the large $\lambda$ strong coupling limit.  
In this limit, the string theory is classical, and the problem of including a D5 brane in the $AdS_5\times S^5$ geometry
reduces to that of finding an extremum of the Dirac-Born-Infeld and Wess-Zumino actions,
\begin{align} \label{dbi5}
S=\frac{ T_5}{g_s} N_5 \int d^{\:6}\sigma\left[- \sqrt{-\det( g+2\pi\alpha'{\mathcal F})} +2\pi\alpha'
C^{(4)}\wedge {\mathcal F}\right], 
\end{align}
where $g_s$ is the closed string coupling constant, which is related to the ${\mathcal N}=4$
Yang-Mills coupling
by $4\pi g_s =g_{YM}^2$, $\sigma^a$ are the coordinates of the D5 brane
worldvolume, $g_{ab}(\sigma)$ is the induced metric of the D5 brane, $C^{(4)}$ is the  4-form of the $AdS_5\times S^5$ background,
${\mathcal F}$ is the worldvolume gauge field and
\begin{align}\label{t5}
T_5=\frac{1}{(2\pi)^5{\alpha'}^3},
\end{align} 
is the D5 brane tension.  Note the overall factor of $N_5$ in equation (\ref{dbi5}). The Wess-Zumino action will not contribute
to the D5 brane equations of motion for the types of embeddings that we will discuss here. 

We shall work with coordinates where  the metric of $AdS_5\times S^5$ is
\begin{align}
ds^2= \sqrt{\lambda}\alpha'&\left[ r^2 (-dt^2+dx^2+dy^2+dz^2) + \frac{dr^2}{r^2}+\right. \nonumber \\
&\left. +d\psi^2+\sin^2\psi(d\theta^2+\sin^2\theta d\phi^2)+ \cos^2\psi (d\tilde\theta^2+\sin^2\tilde\theta d\tilde\phi^2)\right].
\label{ads5metric}\end{align}
Here, $(t,x,y,z,r)$ are coordinates of the Poincare patch of $AdS_5$.  
The boundary of $AdS_5$ is located
at $r\to\infty$ and the Poincare horizon at $r\to 0$.  The coordinates of $S^5$ are a fibration of the 5-sphere by two 2-spheres
over the interval $\psi\in[0,\pi/2]$.

The dynamical variables are
the ten functions of six worldvolume coordinates which embed the D5 brane in $AdS_5\times S^5$,
as well as the six worldvolume gauge fields. 
We shall take the  static gauge where the D5 brane has   coordinates
$(t,x,y,r,\theta,\phi)$ and, due to symmetry,  the remaining two coordinates 
$z(r)$ and $\psi(r)$ depend only on the AdS radius $r$.  The equation of motion for $z$ is satisfied
by a constant.   The  worldvolume metric is
\begin{align}\label{D5metric}
ds^2= \sqrt{\lambda}\alpha'\left[ r^2 (-dt^2+dx^2+dy^2)+\frac{dr^2}{r^2}\left(  1+\left( r\frac{d\psi}{dr} \right)^2 \right)
+\sin^2\psi(d \theta^2+\sin^2 \theta d\phi^2)\right].
\end{align} 
If $r\tfrac{d}{dr}\psi(r)\to 0$ at $r\to\infty$, in the large $r$ regime, the worldvolume is a product of  
$AdS_4$ and $ S^2$. 
We shall also take the worldvolume gauge fields with field strength
\begin{align}
2\pi\alpha'{\mathcal F}=\sqrt{\lambda}\alpha'\left[ \frac{d}{dr}a(r)dr\wedge dt + bdx\wedge dy \right].
\end{align}
The constant external magnetic field is
\begin{align}\label{externalmagneticfield}
B= \frac{\sqrt{\lambda}}{2\pi}~b,
\end{align}
and the r-dependent temporal component of the gauge field is
\begin{align}\label{gaugefield}
A_t(r)= \frac{\sqrt{\lambda}}{2\pi}~a(r).
\end{align}
With the above ansatz for the worldvolume metric and gauge fields, the Born-Infeld action becomes
\begin{align}\label{ansatzaction5}
S_5=-{\mathcal N}_5N_5\int_0^\infty  dr~  2\sin^2\psi\sqrt{b^2+r^4}\sqrt{1+\left(r\frac{d\psi}{dr}\right)^2-\left(\frac{da}{dr}\right)^2},
\end{align}
  where, using (\ref{t5}),
\begin{align}\label{N5}
{\mathcal N}_5=\frac{T_5}{g_s} (\sqrt{\lambda}\alpha')^3(2\pi)V_{2+1}=\frac{2\sqrt{\lambda}  N}{(2\pi)^3}V_{2+1}.
\end{align}
The factor of  $\frac{T_5}{g_s}$ in (\ref{N5}) is the coefficient of the Dirac-Born-Infeld action (\ref{dbi5}), 
$T_5$ is given in (\ref{t5}) and we recall that $g_s=\lambda/(4\pi N)$. The factor
$ (\sqrt{\lambda}\alpha')^3$  comes from the overall factor in the worldvolume metric
in equation (\ref{D5metric}).  The factor of $(2\pi)$ is from the integral over the worldvolume two-sphere.
\footnote{It is  half of the volume of the 2-sphere.  The other factor of 2 is still in the
action in front of $\sin^2\psi$. This notation is designed to match with the D7 brane, which we shall study
in the next section, and to coincide with notation in reference \cite{Bergman:2010gm}.}  
The integral over $(x,y,t)$ produces the volume factor $V_{2+1}$.

Now, we must solve the equations of motion for the functions $\psi(r)$ and $a(r)$  which result
from the Lagrangian (\ref{ansatzaction5}) and the variational principle. 
We will use the boundary condition
\begin{align}
\lim_{r\to\infty}\psi(r)=\frac{\pi}{2},
\end{align}
which is compatible with the equation of motion for $\psi$.   
 Since the Lagrangian depends only
on the derivative of $a(r)$ and not on the variable  $a(r)$ itself, $a(r)$ is 
a cyclic variable and  can be eliminated using its  equation  of motion,
\begin{align} \label{const25}
\frac{d}{dr}\frac{\delta S_5}{\delta \frac{d}{dr}a(r)}=0~~\to~~
  \frac{2\sin^2\psi\sqrt{(b^2+r^4)} {\frac{d}{dr}a} }{\sqrt{1+\left(r\frac{d\psi}{dr}\right)^2-\left(\frac{da}{dr}\right)^2} } =q_5,
\end{align}
where $q_5$ is a constant of integration.  
 We can solve for $\frac{d}{dr}a$,
\begin{align}
\frac{d}{dr}a=\frac{q_5\sqrt{1+\left(r\frac{d\psi}{dr}\right)^2}}{\sqrt{4\sin^4\psi (b^2+r^4)+q_5^2}} \label{aprime5}.
\end{align}
We now must solve the remaining problem of determining $\psi(r)$ where we wish to extremize the Dirac-Born-Infeld action
for a fixed value of the integration constant $q_5$ which will be proportional  to the total electric charge. To this end, we
use the Legendre transformation
$$
{\mathcal R}_5= S_5-\int dr\frac{d}{dr}a(r)\frac{\partial L_5}{\partial \frac{d}{dr}a(r)},
$$
to eliminate $\frac{d}{dr}a$.  We obtain the Routhian
\begin{align}
{\mathcal R}_5=-{\mathcal N}_5N_5\int d r  \sqrt{4\sin^4\psi (b^2+r^4)+ q_5^2}\sqrt{1+\left(r\frac{d\psi}{dr}\right)^2},
\label{routhian5}
\end{align}
which must now be used to find an equation for  $\psi(r)$.  Applying the Euler-Lagrange equation to the Routhian
(\ref{routhian5})  leads to 
\begin{align}
\frac{\left(r\frac{d}{dr}\right)^2\psi}{1+\left({r\frac{d}{dr}\psi}\right)^2}+r\frac{d}{dr}\psi\left[ 1+\frac{8r^4\sin^4\psi   }{4\sin^4\psi
  (b^2+r^4)+q_5^2}\right]
-\frac{ 8\sin^3\psi\cos\psi  (b^2+r^4) }{4\sin^4\psi (b^2+r^4) +q_5^2}=0.
 \nonumber \end{align}
Remembering that the total charge density is defined by 
 $$
\rho = \frac{1}{V_{2+1}}\frac{\delta S_5}{\delta \frac{d}{dr}A_t(r)}=\frac{1}{V_{2+1}}\frac{2\pi}{\sqrt{\lambda}}
\frac{\delta S_5}{\delta \frac{d}{dr}a(r)}=\frac{1}{V_{2+1}}\frac{2\pi}{\sqrt{\lambda}}{\mathcal N}_5N_5q_5 
= \frac{2 NN_5}{(2\pi)^2}~q_5,
$$
we see that $q_5$ is related to the total charge density $\rho$ by
\begin{align} \label{const35}
q_5=\frac{(2\pi)^2}{2 NN_5}~\rho.   
\end{align}
Without loss of generality, we can rescale $r$ in the equation of motion  so that $b=1$.  
Then, $q_5$ in that equation is replaced by $q_5/b$, which, for future reference we write as
\begin{align}\label{d5100}
\frac{q_5}{b}= \frac{\pi\nu}{f},
\end{align}
where the parameter $\nu$ defined in (\ref{definenu}) 
is the filling fraction of a free particle Landau level (the density of a Landau level is $B/2\pi$), normalized by a factor
of the number of colors of the bifundamental charged fields. If $N$ of the fundamental ``quark' fields must bind to form a baryon,
which has U(1) charge $N$, then $\nu$ would be the filling fraction of a single Landau level of noninteracting baryons.
The factor in the denominator of (\ref{d5100}) is related
to the total number of D5 branes, 
\begin{align}
f\equiv\frac{2\pi}{\sqrt{\lambda} } N_5.
\label{definef}
\end{align}
This will be an important parameter in the next section when we consider the D7 branes.   Here we note that we 
are interested in the parameter regime where $f$ is of order one when $\lambda$ is large.  This means that
the number of D5 branes, $N_5=\frac{\sqrt{\lambda}}{2\pi}f$ is large. Finally, we present the equation of motion as
\begin{align}
\frac{\left(r\frac{d}{dr}\right)^2\psi}{1+\left({r\frac{d}{dr}\psi}\right)^2}+r\frac{d}{dr}\psi\left[ 1+\frac{8r^4\sin^4\psi  f^2 }{4\sin^4\psi
  f^2(1+r^4)+(\pi\nu)^2}\right]  \nonumber \\
-\frac{ 8\sin^3\psi\cos\psi f^2 (1+r^4) }{4\sin^4\psi f^2(1+r^4) +
(\pi \nu)^2}=0.
\label{equationforpsi5}\end{align}
The Routhian from which this equation is derived is
\begin{align}
{\mathcal R}_5=-\frac{{\mathcal N}_5N_5}{f}\left(\frac{2\pi B}{\sqrt{\lambda}} \right)^{\frac{3}{2}}\int_0^\infty d r  ~\sqrt{4\sin^4\psi f^2(1+r^4)+ (\pi\nu)^2}\sqrt{1+\left(r\frac{d\psi}{dr}\right)^2}.
\label{routhian51}
\end{align}
We can now see that the boundary condition that $\psi\to\frac{\pi}{2}$ at $r\to\infty$ is compatible with the equation of motion.  In fact, $\psi=\frac{\pi}{2}$ is a solution of the equation of motion for all values of $r$.   However, this constant solution  
is known to be unstable if the filling fraction is small enough \cite{Jensen:2010ga}.
The critical value of the filling fraction is 
\begin{equation}
\nu_c = 2\sqrt{7} f/\pi\sim 1.68  f.
\label{criticalnu}
\end{equation}
When $|\nu|<\nu_c$, the stable solution of the equation of motion has r-dependent $\psi$. 
When $|\nu|>\nu_c$, constant $\psi=\frac{\pi}{2}$ is the stable solution.
Looking at the equation for $\psi(r)$ in the large $r$ regime, 
 \begin{align}
\left(r\frac{d}{dr}\right)^2\delta \psi + 3\left(r\frac{d}{dr}\right)\delta \psi + 2\delta\psi=0,
\end{align}
we see that it must have the asymptotic behavior
\begin{align}
\psi(r)=\frac{\pi}{2} + \frac{c_1}{r}+\frac{c_2}{r^2}+\ldots
\label{asymptoticpsi}
\end{align}
The constant $c_1$ has the holographic interpretation of being proportional to the bare mass of the fundamental representation fields\footnote{ The
constant $c_1$ can indeed 
be seen to be proportional to the asymptotic at large $r$ 
physical separation of the D5 and D3 branes.  When this separation is non-zero, the open strings connecting the
D3 and D5 branes have a minimum energy, or a mass gap. This is interpreted as a bare mass for the hypermultiplet fields of the field theory dual.  Note 
that such a mass breaks and $\tilde{SO(3)}$ symmetry explicitly, by separating the D3 and D5 branes in one of the transverse directions.  Since the 
$\tilde{SO(3)}$ rotates
the transverse directions into each other, specifying the direction of  separation in those directions breaks the  symmetry. The field theory
operator dual to fluctuations of this separation is a parity symmetric, $\tilde{SO(3)}$ breaking mass term for the fermions in the hypermultiplet.  The
exponents $1$ and $2$ of $\frac{1}{r}$ and $\frac{1}{r^2}$ in (\ref{asymptoticpsi}) correspond to conformal dimensions of the mass and mass operator which, since 
they are protected by supersymmetry in the D3-D5 system, coincide with their tree level, engineering dimensions.}.  
The constant $c_2$ is proportional to the chiral condensate.

One of these constants is to be specified in the boundary conditions which define the theory.  
The other constant is then determined by solving the equations of motion.  In this Paper,
we will restrict our attention to the massless theory, that is, we will set $c_1=0$ and and, consequently, 
$c_2$ will be determined by the dynamics.   

Then, if it happens that $c_2=0$, because $\psi$ must satisfy the second order differential equation (\ref{equationforpsi5}), it must be a constant , $\psi=\frac{\pi}{2}$ for all values of $r$.  
This is the solution with vanishing chiral condensate which is
stable when the charge density is big enough,  $|\nu|>\nu_c$.  

If $c_2\neq 0$, the chiral condensate is nonzero and $\psi$ is r-dependent.  Since the bare mass is zero, the chiral condensate being nonzero is
interpreted as spontaneous breaking of chiral symmetry.  This phase is stable when $|\nu|<\nu_c$.  

The phase transition which occurs when $|\nu|=\nu_c$
is thus interpreted as a chiral symmetry breaking/restoring phase transition which is driven by decreasing/increasing the quark density. 
This phase transition was found
in reference \cite{Jensen:2010ga}.  One very interesting feature is that it has Berezinsky-Kosterlitz-Thouless scaling
and it is thus one of the few holographic phase transitions that does not exhibit mean field scaling. 

We also note that, when $\nu\neq 0$, even when chiral symmetry is broken and the quarks have dynamically generated masses, the
finite density system that we find is compressible.   As we have already discussed, the D5 brane must necessarily reach the Poincare horizon. 
One way to understand this from the Routhian (\ref{routhian51}) is to note that, if the brane were to cap off at a finite radius $r_0$, the
derivative $d\psi/dr$ would diverge there.  This would result in a very large energy unless the first square root factor in the Routhian compensated
it by going to zero.  This cannot happen when $\nu\neq 0$, since that factor is bounded from below by $|\pi\nu|$.  
In the neutral case, when $\nu=0$, it is allowed, $\psi\to 0$ as $r\to r_0$ and 
the square root factor would indeed vanish in that special case and it can thereby keep the energy finite.

 In the next section, we shall derive the equations of motion for the D7 brane with the same charge density, magnetic field and 
asymptotic boundary condition (\ref{asymptoticpsi}) where we will also set $c_1=0$.

\section{D3-D7}

We shall study the D3-D7 system in the same probe limit in which we 
considered the D5 branes in the previous section.  On flat space, the D3, D5 and D7 branes are oriented
as in Table 1. In this case, there
will be a single D7 brane whose worldsheet will wrap both of the spheres
$S^2$ and $\tilde S^2$.    The worldsheet gauge fields have
$N_5$ units of magnetic flux on $\tilde S^2$.   The Dirac-Born-Infeld and Wess-Zumino actions for the D7 brane are
\begin{align} 
S=\frac{ T_7}{g_s} \int d^{\:8}\sigma\left[- \sqrt{-\det(   g+2\pi\alpha' {\mathcal F})} +\frac{(2\pi\alpha')^2}{2}
C^{(4)}\wedge  {\mathcal F}\wedge  {\mathcal F}\right], 
\end{align}
where 
\begin{align}\label{t7}
T_7=\frac{1}{(2\pi)^7{\alpha'}^4},
\end{align} 
is the D7 brane tension.  

We shall work with coordinates where  the metric of $AdS_5\times S^5$ are given in (\ref{ads5metric}) and 
the Ramond-Ramond  4-form  of the IIB supergravity background in this coordinate system is
\begin{align}\label{4form}
C^{(4)}=  \lambda{\alpha'}^2\left[
r^4dt\wedge dx\wedge dy\wedge dz+ \frac{c(\psi)}{2}d\cos\theta\wedge d\phi
\wedge d\cos\tilde\theta\wedge d\tilde\phi\right],
\end{align}
Here, 
$
\partial_\psi c(\psi)=8\sin^2\psi\cos^2\psi
$
which has the indefinite integral
\begin{align}
c(\psi)= \psi - \frac{1}{4}\sin4\psi -\frac{\pi}{2}.
\end{align}
Here we have chosen the integration constant so that $c(\pi/2)=0$, i.e.\
it vanishes at the asymptotic value of the angle $\psi$.
The choice of this constant is a string theory gauge choice and our results will not depend on its specific value. 

In  the ``static gauge''  the D7-brane has   coordinates
$(t,x,y,r,\theta,\phi,\tilde\theta,\tilde\phi)$.  The equation of motion for $z$ is satisfied
by a constant.   The D7 brane worldvolume metric is
\begin{align}\label{D7metric}
ds^2= \sqrt{\lambda}\alpha'\left[ r^2 (-dt^2+dx^2+dy^2)+\frac{dr^2}{r^2}\left(1+\left(r\frac{d\psi}{dr}\right)^2\right)+\right. \nonumber \\ \left. +\sin^2\psi(d\theta^2+\sin^2\theta d\phi^2) +\cos^2\psi(d\tilde\theta^2+\sin^2\tilde\theta d\tilde\phi^2)\right],
\end{align} 
The worldvolume gauge fields are
\begin{align}
2\pi\alpha'{\mathcal F}=\sqrt{\lambda}\alpha'\left( \frac{d}{dr}a(r)dr\wedge dt + bdx\wedge dy + \frac{f}{2}d\cos\tilde\theta\wedge d\tilde\phi\right).
\end{align}
The parameter $f $ is the  flux of the worldvolume gauge fields defined in equation (\ref{definef}). 
It corresponds to $N_5$
Dirac monopoles on $\tilde S^2$.
  The external magnetic field is identical to that on the D5 brane. 
The temporal component of the gauge field  will be determined so that the total charge density is $\rho$.
Substituting the ansatz into the Dirac-Born-Infeld action, we obtain
\begin{align}\label{ansatzaction}
S_7=-{\mathcal N}_7\int_0^\infty  dr\left[2\sin^2\psi\sqrt{(f^2+4\cos^4\psi)(b^2+r^4)}
\sqrt{1+\left(r\frac{d\psi}{dr}\right)^2-\left(\frac{da}{dr}\right)^2} \right. 
\nonumber \\ \left. +  2\,
\frac{da}{dr} b c(\psi)\right],
\end{align}
  where, using (\ref{t7}),
\begin{align}\label{N}
{\mathcal N}_7= \frac{2\lambda  N}{(2\pi)^4}V_{2+1}.
\end{align}
 We shall use the same boundary condition for $\psi(r)$ that we imposed in the case of the D5 brane in (\ref{asymptoticpsi}). 
This boundary condition is compatible with the equation of motion for $\psi(r)$ which we shall derive below.  

Now, in a completely analogous way to the D5 brane case, 
we must solve the equations of motion for the functions $\psi(r)$ and $a(r)$  which result
from the action (\ref{ansatzaction}) and the variational principle. 
Again,  $a(r)$ is 
a cyclic variable and  can be eliminated using its  equation  of motion,
\begin{align} \label{const27}
\frac{d}{dr}\frac{\delta S_7}{\delta \frac{d}{dr}a(r)}=0~~\to~~
  \frac{2\sin^2\psi\sqrt{(f^2+4\cos^4\psi)(b^2+r^4)} {\frac{d}{dr}a} }{\sqrt{1+\left(r\frac{d\psi}{dr}\right)^2-\left(\frac{da}{dr}\right)^2 } }
-2bc=q_7,
\end{align}
where $q_7$ is a constant of integration.  
It is proportional to the total charge density
in the field theory dual
 \begin{align} \label{const2}
\rho = \frac{1}{V_{2+1}}\frac{\delta S_7}{\delta \frac{d}{dr}A_t(r)}=\frac{1}{V_{2+1}}\frac{2\pi}{\sqrt{\lambda}}{\mathcal N_7}q_7 
~~,~~   q_7 = \frac{(2\pi)^3}{2\sqrt{\lambda}N}\rho.
\end{align}
 We can solve for $\frac{d}{dr}a$,
\begin{align}
\frac{d}{dr}a=\frac{(2bc+q_7)\sqrt{1+\left(r\frac{d\psi}{dr}\right)^2}}{\sqrt{4\sin^4\psi(f^2+4\cos^4\psi) (b^2+r^4)+(q_7+2bc)^2}}. \label{aprime}
\end{align}
The Legendre transformation
yields the Routhian
\begin{align}
{\mathcal R}_7=-{\mathcal N_7}\int_0^\infty d r  \sqrt{4\sin^4\psi(f^2+4\cos^4\psi) (b^2+r^4)+(q_7+2bc)^2}\sqrt{1+\left(r\frac{d\psi}{dr}\right)^2},
\label{routhian1}
\end{align}
which must now be used to find an equation of motion for  $\psi(r)$.  Applying the Euler-Lagrange equation to the Routhian
(\ref{routhian1})  leads to 
\begin{align}
\frac{\left(r\frac{d}{dr}\right)^2\psi}{1+\left(r\frac{d\psi}{dr}\right)^2}+r\frac{d\psi}{dr}\left[ 1+\frac{8r^4\sin^4\psi (f^2+4\cos^4\psi) }{4\sin^4\psi
(f^2+4\cos^4\psi) (1+r^4)+(\pi(\nu-1) +2\psi-\frac{1}{2}\sin4\psi)^2}\right] + \nonumber \\
-\frac{ 8\sin^3\psi\cos\psi f^2(1+r^4) + 4\sin^32\psi\cos2\psi r^4 
+4\sin^22\psi(\pi(\nu-1)+2\psi)
}{4\sin^4\psi(f^2+4\cos^4\psi) (1+r^4) +(\pi(\nu-1) +2\psi-\frac{1}{2}\sin4\psi)^2}=0.
\label{equationforpsi}\end{align} 
We have rescaled $r$ so that $b=1$.  Here
$$
\frac{q}{b}=  \frac{4\pi^3}{\sqrt{\lambda}  N}\rho \cdot \frac{\lambda}{2\pi}\frac{1}{B}
= \pi \nu,
$$
where $\nu$ is defined in equation (\ref{definenu}). The Routhian is
\begin{align}\label{routhian720}
{\mathcal R}_7&=-{\mathcal N_7}\left(\frac{2\pi B}{\sqrt{\lambda}}\right)^{\frac{3}{2}}\int_0^\infty d r  L_7,
\end{align}
where
\begin{align}
L_7=\sqrt{4\sin^4\psi(f^2+4\cos^4\psi) (1+r^4)+(\pi(\nu-1) +2\psi-\frac{1}{2}\sin4\psi)^2}
\sqrt{1+\left(r\frac{d\psi}{dr}\right)^2}.
\label{routhian72}
\end{align}
Now, we note the similarity of the Routhian (\ref{routhian72}) for the D7 brane to that of the D5 brane (\ref{routhian51}). 
At the asymptotic boundary, $r\to\infty$, the Lagrangians become identical. To see this, we note 
the identity
\begin{align}\label{normalizationidentity}
{\mathcal N}_7f = {\mathcal N}_5 N_5.
\end{align}

We can now see that, from the form of the Routhian, the D7 brane can have a Minkowski embedding when 
the first square root in $L_7$ in (\ref{routhian72}) can go to zero in order to compensate the fact that at a smooth
cap-off, $d\psi/dr$ must diverge and make the second factor large.  The first square root contains a sum of 
squares which can go to zero only 
when  $\psi$ goes to zero and 
only when $\nu=1$, the charge density coinciding with the first integer Hall filling fraction. 

We note that the $r$-independent chirally symmetric solution, $\psi=\frac{\pi}{2}$ is also a solution of the D7 brane
equation of motion (\ref{equationforpsi}). We would expect that this  solution is stable if the charge density $\nu$ is
large enough.  In fact, by linearizing the equation about the constant solution, we can see that it is stable to small fluctuations
in precisely the same region that the D5 brane is stable, for $|\nu|>\nu_c$. When $|\nu|<\nu_c$, like the D5 brane, the D7
brane must have an $r$-dependent solution which breaks chiral symmetry. Our numerical results will indicate that there are
$r$-dependend D7 solutions even in the region where the constant solution is stable to small fluctuations, and these solutions
have lower energies than the constant solution.  This means that the chiral symmetry breaking solution is actually the prefered one
there, and that there is an energy barrier between the two. 

Now, we can compare the mathematical problems of finding the D5 and D7 brane embedding with the same 
values of external parameters, $B$, $\nu$ and $c_1$, the latter being set to zero.

  \section{Numerical Solutions}

We must solve the equations (\ref{equationforpsi5}) and (\ref{equationforpsi}) with the boundary condition (\ref{asymptoticpsi})
where we set $c_1=0$. 
For each differential equation we start by establishing a high order power
series solution, valid at large $r$, which obeys the boundary 
condition and 
has $c_2$, $f$ and $\nu$ as parameters.  
With the aid of these series solutions 
we then generate initial data for $\psi(r)$ and 
$r\frac{d}{dr}\psi(r)$, at a large value of $r$, which we use 
as input to a numerical integration routine in Mathematica that allows us
to integrate the
prospective solution to smaller values of $r$. In this process
we use a shooting technique which, for a given set of values of $(\nu,f)$
enables us to determine the value of $c_2$ for which
a satisfactory solution exists.   This value of $c_2$ is interpreted as the chiral condensate.   
The numerical  value of  $c_2$ itself is different for the D5 and the D7 solution.   

For the solutions we have found numerically we can then calculate 
the energy densities which (for the static configurations
that we have been considering) are proportional to the negatives of the 
Routhians (\ref{routhian51}), (\ref{routhian720})  and (\ref{routhian72}).  
Finally, by determining the total energy of the D5 and D7 brane solutions
corresponding to a given value of $f$ and $\nu$ we can determine which
configuration is the stable one.\footnote{When integrating the energy
densities we use the earlier mentioned series expansions to estimate
the contributions from the region of very large $r$ and we approximate
the densities with their asymptotic constant values to estimate the contribution
from the small-$r$ region.}

\begin{figure}
\begin{center}
\includegraphics[scale=0.7]{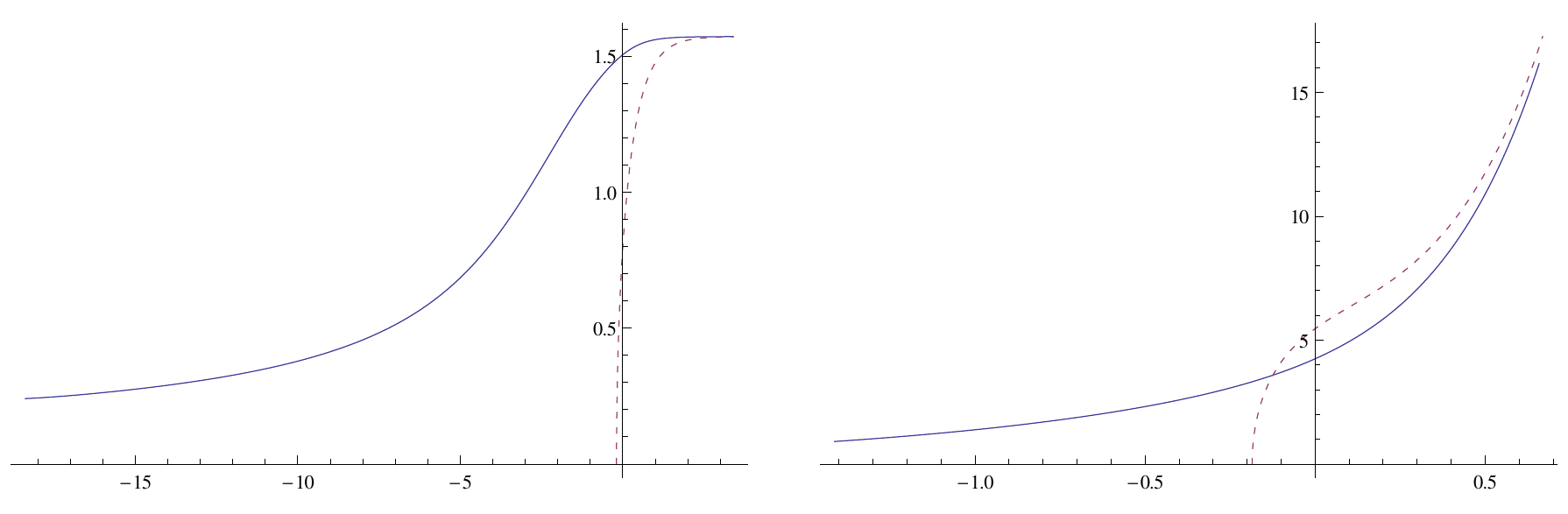}
\end{center}
\caption{\label{fig5} {\footnotesize Numerical solution with filling fraction $\nu=1$ and $f=1$. 
  On the left-hand-graph, 
the profile of the function $\psi$ is plotted on the vertical 
axis versus $\ln(r)$ on the horizontal axis.
 The  D7 brane is the dashed curve which comes to the horizontal axis.
The D5 brane is the solid curve approaching the the horizontal axis at the Poincare horizon 
which is at the asymptotic left-hand-side 
of the graph.
 The second graph is the energy density on the vertical axis versus $\ln(r)$ on
the horizontal axis.  It should be integrated
over $\ln(r)$ to get the total energy. 
Numerical computation shows that the
D7 brane has lower energy,  $E5-E7=1.71
\cdot\frac{2  N}{(2\pi)^{\frac{5}{2}}\sqrt{\lambda}} B^{\frac{3}{2}}$, and is therefore the preferred state.}}
\end{figure}

\begin{figure}
\begin{center}
\includegraphics[scale=.7]{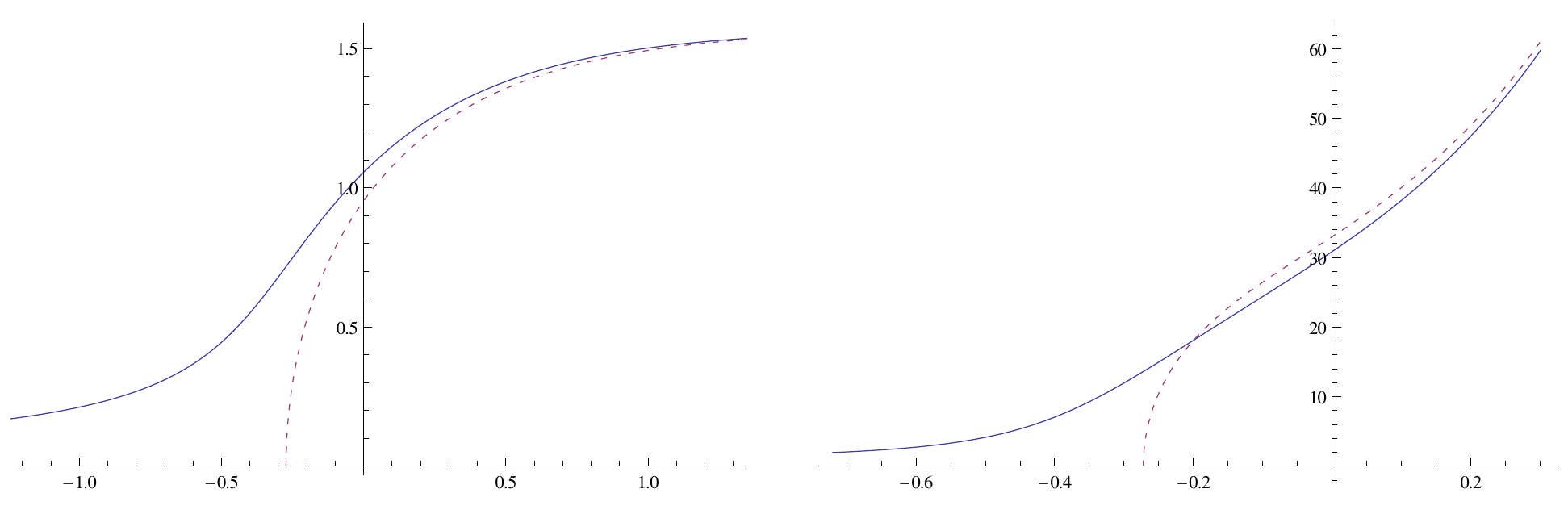}
\end{center}
\caption{\label{fig6} {\footnotesize Numerical solution with filling fraction $\nu=1$ and $f=10$. 
 On the left-hand-graph, 
the profile of the function $\psi$ is plotted on the vertical 
axis versus $\ln(r)$ on the horizontal axis.
 The  D7 brane is the dashed curve and the  D5 brane is the solid curve.  
 The second graph compares the energy densities.  
The D7 brane is the stable solution. 
Numerically, $E5-E7=2.34
\cdot\frac{2  N}{(2\pi)^{\frac{5}{2}}\sqrt{\lambda}} B^{\frac{3}{2}}$. }}
\end{figure}

 Numerical solutions
for the profile of $\psi(r)$ for the D5 and D7 branes when $\nu=1$ and when $f=1$ and $f=10$ are plotted in figures \ref{fig5} and \ref{fig6}
respectively.   The case $\nu=1$ is the one where the D7 brane is allowed to have a Minkowski embedding which is clearly seen in the numerical 
plot.  For both of those cases, the D7 brane has lower energy than the D5 brane and it is therefore the preferred solution.

\begin{figure}
\begin{center}
\includegraphics[scale=.7]{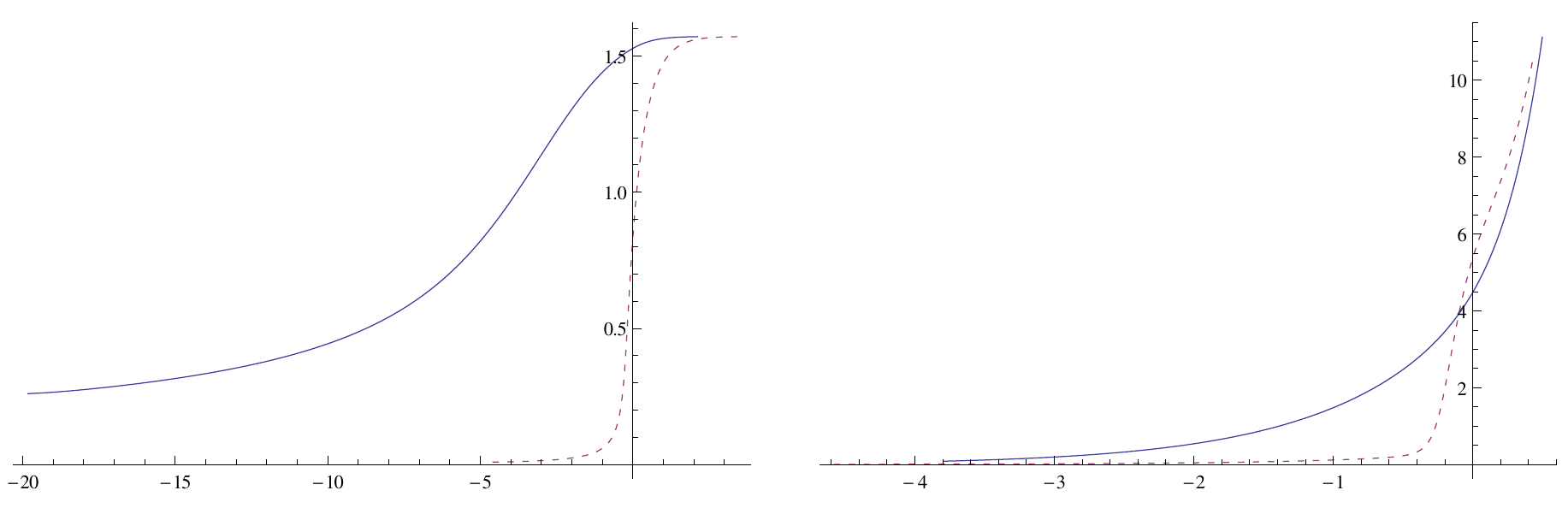}
\end{center}
\caption{\label{fig8} {\footnotesize Numerical solution with filling fraction  $\nu=1.1$ and $f=1$.  On the left-hand-graph, 
the profile of the function $\psi$ is plotted on the vertical 
axis versus $\ln(r)$ on the horizontal axis. The second graph is the energy density on the vertical axis versus $\ln(r)$ in
the horizontal axis. Like the D5 brane, in this case the D7 brane also approaches the Poincare horizon at the asymptotic left-hand-side
of the graph. The D7 brane has lower energy. 
Numerically, $E5-E7=1.66
\cdot\frac{2  N}{(2\pi)^{\frac{5}{2}}\sqrt{\lambda}} B^{\frac{3}{2}}$
}}
\end{figure}

\begin{figure}
\begin{center}
\includegraphics[scale=.7]{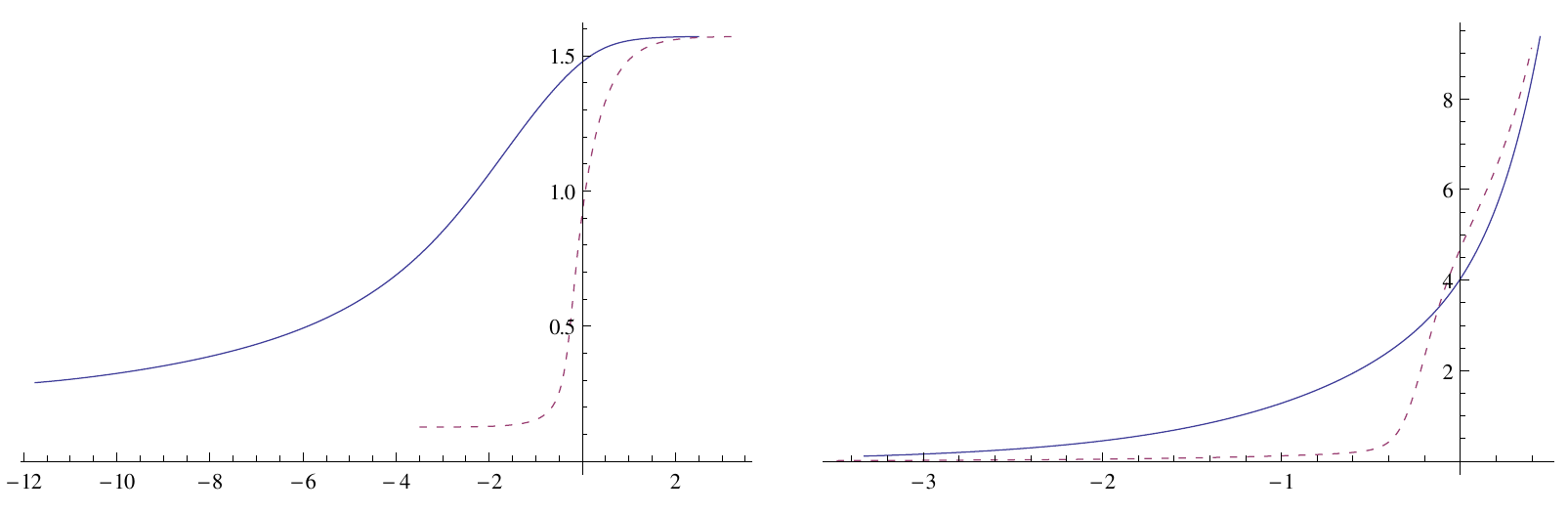}
\end{center}
\caption{\label{fig9} {\footnotesize Numerical solution with filling fraction  $\nu=0.9$ and $f=1$.  On the left-hand-graph, 
the profile of the function $\psi$ is plotted on the vertical 
axis versus $\ln(r)$ on the horizontal axis. The second graph is the energy density on the vertical axis versus $\ln(r)$ in
the horizontal axis.  It should be integrated
over $\ln(r)$ to get the total energy. The D7 brane has lower energy.
Numerically, $E5-E7=1.43
\cdot\frac{2  N}{(2\pi)^{\frac{5}{2}}\sqrt{\lambda}} B^{\frac{3}{2}}$.
}}
\end{figure}

\begin{figure}
\begin{center}
\includegraphics[scale=.7]{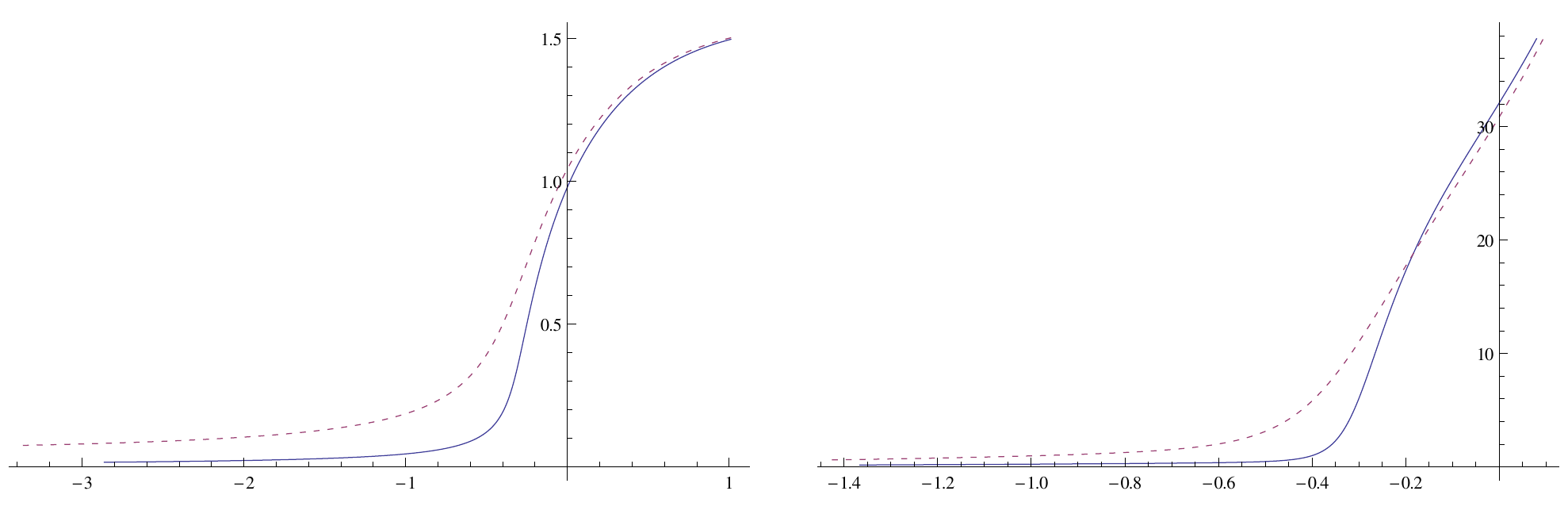}
\end{center}
\caption{\label{fig11} {\footnotesize Numerical solution with filling fraction  $\nu=0.2$ and $f=10$.  On the left-hand-graph, 
the profile of the function $\psi$ is plotted on the vertical 
axis versus $\ln(r)$ on the horizontal axis. The second graph is the energy density on the vertical axis versus $\ln(r)$ in
the horizontal axis.    In this case, the D5 brane has lower energy and is the stable solution.
Numerically, $E5-E7=-1.34
\cdot\frac{2  N}{(2\pi)^{\frac{5}{2}}\sqrt{\lambda}} B^{\frac{3}{2}}$
}}
\end{figure}

\begin{figure}
\begin{center}
\includegraphics[scale=.7]{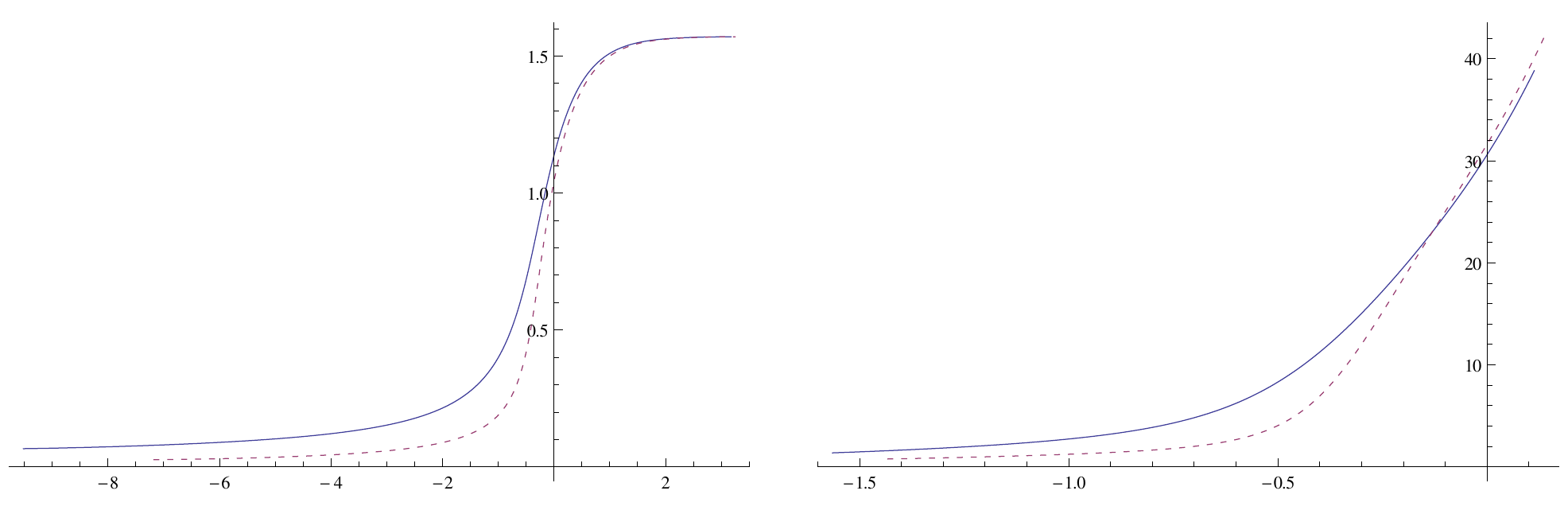}
\end{center}
\caption{\label{fig7} {\footnotesize Numerical solution with filling fraction $\nu=2$ and $f=10$.  On the left-hand-graph, 
the profile of the function $\psi$ is plotted on the vertical 
axis versus $\ln(r)$ on the horizontal axis. The second graph is the energy density on the vertical axis versus $\ln(r)$ in
the horizontal axis.  It should be integrated
over $\ln(r)$ to get the total energy. The D7 brane has lower energy and is the stable solution. Numerically, $E5-E7=2.10
\cdot\frac{2  N}{(2\pi)^{\frac{5}{2}}\sqrt{\lambda}} B^{\frac{3}{2}}$. 
}}
\end{figure}

\begin{figure}
\begin{center}
\includegraphics[scale=.7]{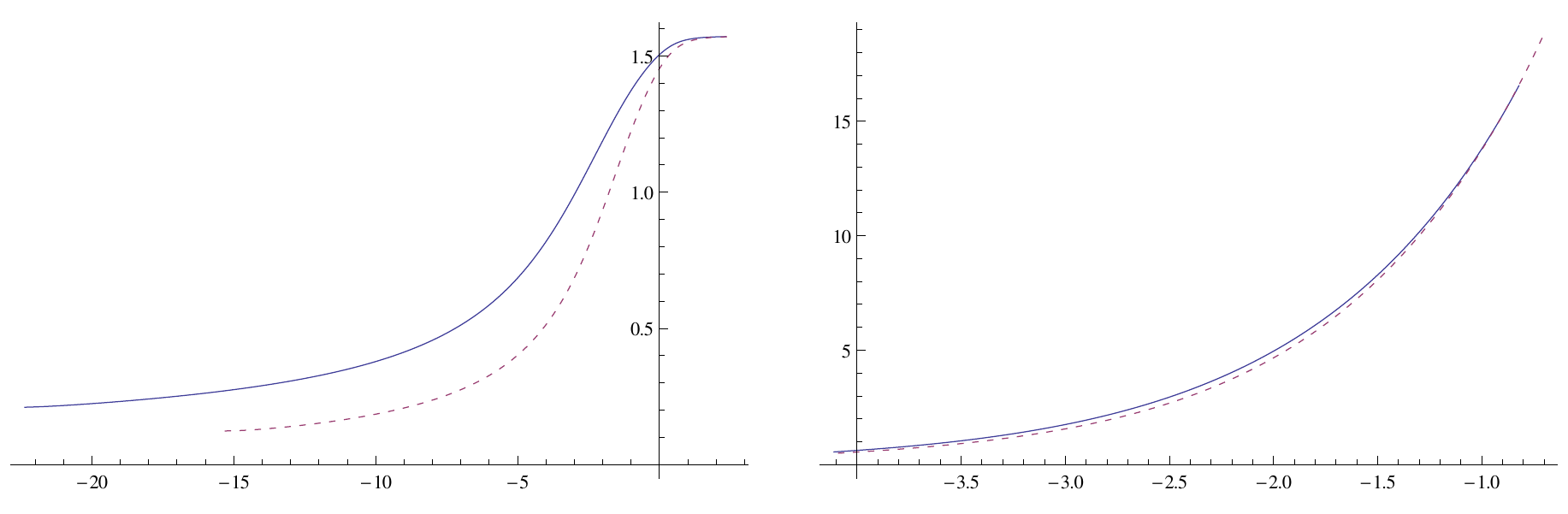}
\end{center}
\caption{\label{fig10} {\footnotesize Numerical solution with filling fraction  $\nu=10$ and $f=10$.  On the left-hand-graph, 
the profile of the function $\psi$ is plotted on the vertical 
axis versus $\ln(r)$ on the horizontal axis. The second graph is the energy density on the vertical axis versus $\ln(r)$ in
the horizontal axis.   The D7 brane has lower energy. Numerically, $E5-E7=0.22
\cdot\frac{2  N}{(2\pi)^{\frac{5}{2}}\sqrt{\lambda}} B^{\frac{3}{2}}$.   
}}
\end{figure}

\begin{figure}
\begin{center}
\includegraphics[scale=.7]{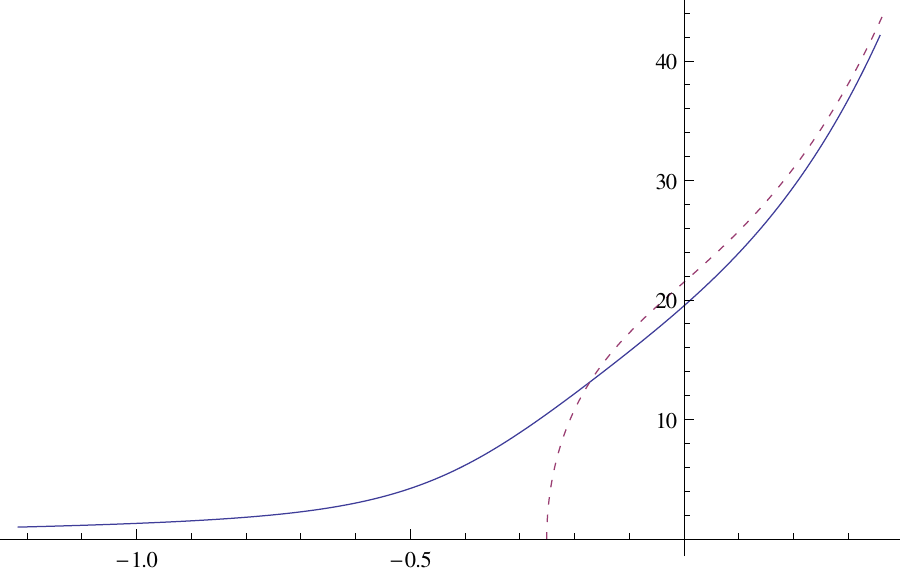}
\end{center}
\caption{\label{fig12} {\footnotesize Numerical solution which compares the energy densities of
a $(\nu=2,f=12)$ state, which has only a black hole embedding, with the energy density of 
two $(\nu=1,f=6)$ states which have Minkowski embeddings.  The vertical axis is energy
density and the horizontal axis is $\ln(r)$.  The energy density (per unit $\ln(r)$) of the $2\times$ $(\nu=1, f=6)$ state
is the dashed curve whereas the energy density of the $(\nu=2,f=12)$
state is the solid curve.  
The sum of the energies of the two $(\nu=1,f=6)$ branes is lower that that
of the $(\nu=2,f=12)$ brane. Numerically,
$E_{(n,f)=(1,12)}-2E_{(n,f)=(1,6)}=2.43 \cdot\frac{2  N}{(2\pi)^{\frac{5}{2}}\sqrt{\lambda}} B^{\frac{3}{2}}$.
}}
\end{figure}

We examine the solutions in the vicinity of $\nu=1$, $(\nu=1.1, f=1)$ and $(\nu=0.9,f=1)$ in figures \ref{fig8} and \ref{fig9}, respectively.
In those cases, like the D5 brane, the D7 brane must have a black hole embedding and this is indeed what is found numerically.  The D7 brane has
a long spike which goes to the Poincare horizon.  Note from the plots that the D7 spike is significantly narrower than the D5 spike.
The D7 brane then loses some of its energy advantage for small 
values of $\ln(r)$.  It nevertheless still has lower total energy than the D5 brane, meaning that even for the non-gapped non-quantum Hall
states, it is the preferred solution.  

The numerical solution for $(\nu=2,f=10)$ is shown in figure \ref{fig7}.  This is an integer filling fraction but the state is not
a quantum Hall state.   We shall explain shortly how to find the quantum Hall state with $\nu=2$ (and which will be the energetically
preferred one).  In the meantime, we note that comparison of energies shows that this intermediate D7  brane state is preferred over the D5 brane.

We have done a numerical exploration of  the region of smaller values of $\nu$.  We find that for $\nu>\frac{1}{2}$, the D7 brane is stable and for $\nu<\frac{1}{2}$ the D5 brane is the stable solution when $f$ is in the region between one and ten.  In figure \ref{fig11} we plot the numerical solution for $(\nu=0.2,f=10)$.
For that solution, we find that the D5 brane is the configuration with lower energy.   
 
Some results for a larger value of $\nu$ are shown in figure \ref{fig7} and \ref{fig10}  where we display the solutions with 
$(\nu=2,f=10)$ and $(\nu=10,f=10)$.  In both of these cases, the D7 brane has lower energy.  We have also looked at an extreme
case where $(\nu=100,f=100)$ and we find that, even though the energies become very close, D7 is still preferred.

Then, we consider some positive integer filling fractions and compare the following solutions, $(\nu=2,f=12)$ and two
D7 branes, each of which have $(\nu=1,f=6)$.  The energy densities of the two cases are plotted in figure \ref{fig11}. 
The energy of the two branes with $\nu=1$ is lower than the single brane with $\nu=2$.  This can be attributed to the fact
that the former has a Minkowski embedding whereas the latter has a black hole embedding.  The Minkowski embedding
has an energy advantage in the small $\ln(r)$ region. 

\begin{figure}
\begin{center}
\includegraphics[scale=.7]{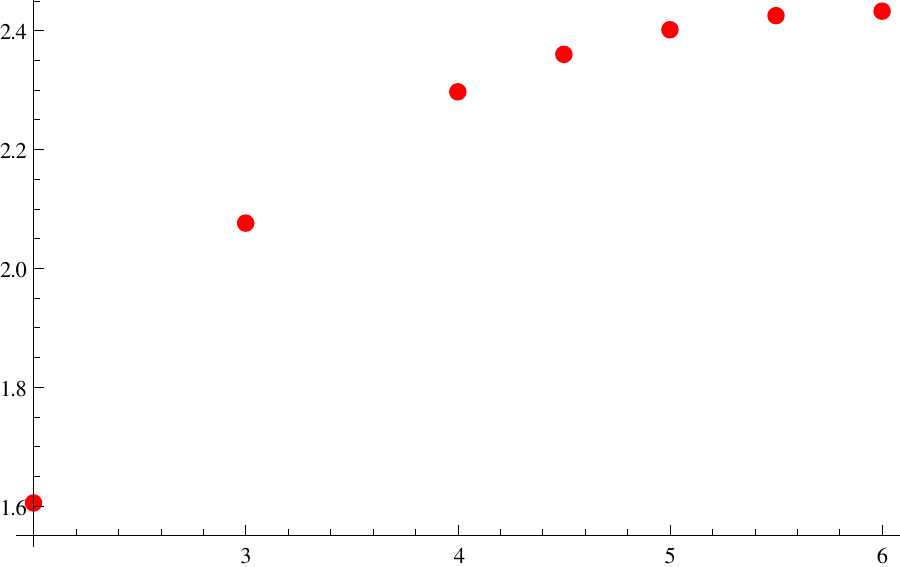}
\end{center}
\caption{\label{fig13} {\footnotesize This is a plot of (a constant minus) the  total energies of two D7 branes, one of
which has $(\nu=1,f_1$ and the other of which has $(\nu=1, 12-f_1)$ for various values of $f_1$ between
0 and 6.    This clearly shows that the symmetric configuration
where $f_1=f/2$ has the lowest energy and is the prefered state. 
}}
\end{figure}

We also consider asymmetric splittings of the $N_5$ D5 branes into two D7 branes when $\nu=2$.  For $f=12$ we consider
a number of cases where the first D7 has $\nu=1$ and $f_1$ and the second D7 has $\nu=1$ and $f_2$ so that $f_1+f_2=f$.
The energies for a number of cases with $f_1$ varying from 2 to $\frac{f}{2}=6$ are plotted in figure \ref{fig13}.

\begin{figure}
\begin{center}
\includegraphics[scale=.5]{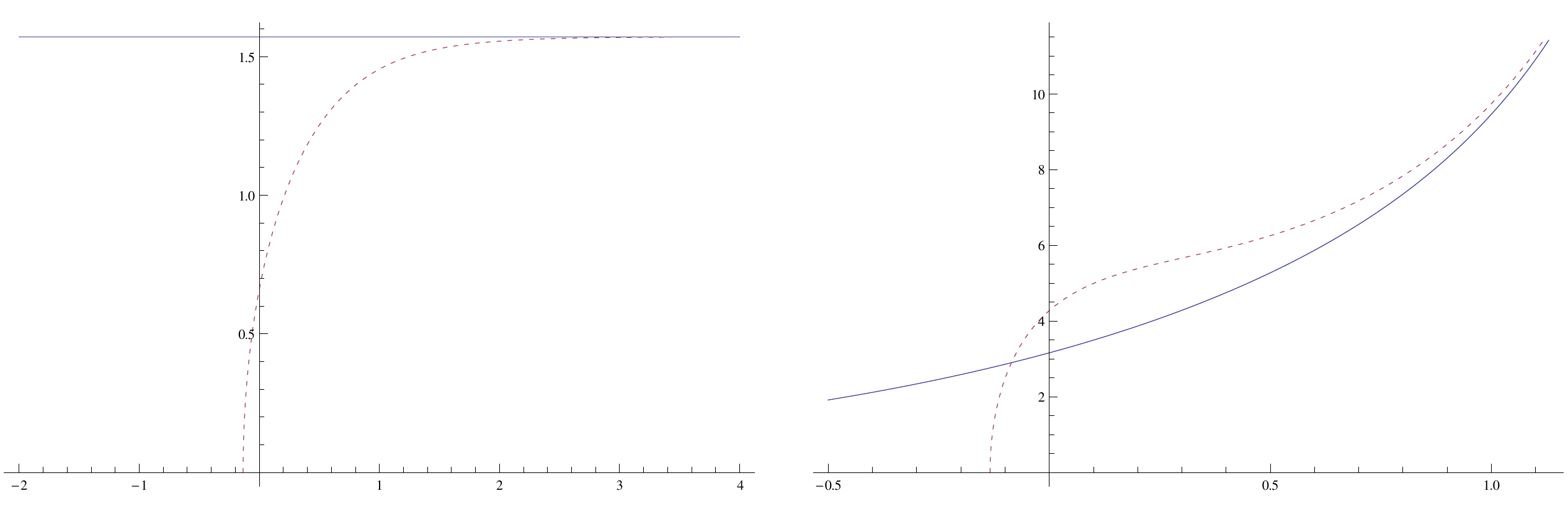}
\end{center}
\caption{\label{fig14} {\footnotesize The solution for a  D7 brane where $\nu=1$ and $f=0.1$  is depicted (dotted lines)
in the left-hand-graph and  its energy density (as a function of $\ln(r)$) in the right-hand graph.  The solid line  in both
graphs is the profile and energy density of the constant $\psi=\frac{\pi}{2}$ solution which is the only
D5 solution  in the $\nu>\nu_c$ regime. 
   For this value of $f$, $\nu_c\approx .168$ and $\nu=1$. 
By integrating the energy density, we clearly see that the chiral symmetry breaking D7 brane
is still energetically preferred, $E_5-E_7= 1.71\cdot\frac{2  N}{(2\pi)^{\frac{5}{2}}\sqrt{\lambda}} B^{\frac{3}{2}}$.
}}
\end{figure}

\begin{figure}
\begin{center}
\includegraphics[scale=.5]{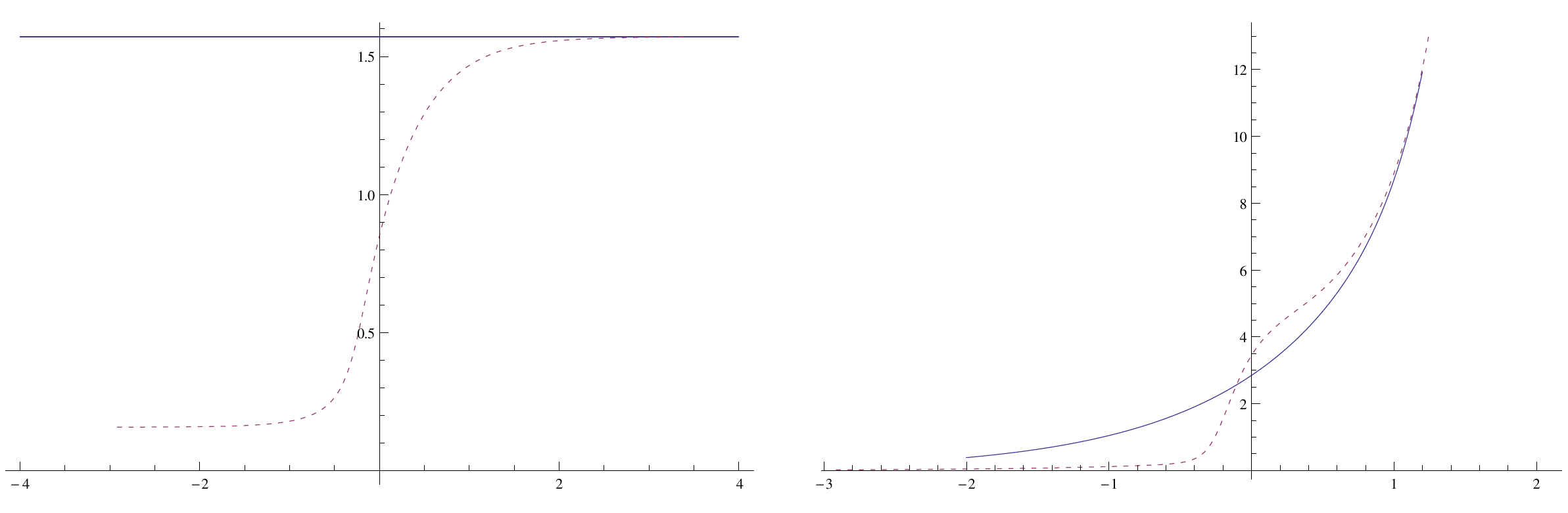}
\end{center}
\caption{\label{fig15} {\footnotesize The solution for a  D7 brane  where $\nu=0.9$ and $f=0.1$ is depicted (dotted lines)
in the left-hand-graph and  its energy density (as a function of $\ln(r)$) in the right-hand graph.  The solid line  in both
graphs is the profile and energy density of the constant $\psi=\frac{\pi}{2}$ solution which is the only 
D5 solution in the $\nu>\nu_c$ regime. 
For this value of $f$, $\nu_c\approx .168$ and $\nu=0.9$. 
By integrating the energy density, we clearly see that the chiral symmetry breaking D7 brane
is still energetically preferred, $E_5-E_7=1.26\cdot\frac{2  N}{(2\pi)^{\frac{5}{2}}\sqrt{\lambda}} B^{\frac{3}{2}}$.
}}
\end{figure}

Finally, we examine the question of whether the chiral symmetry restoring phase transition which occurs for the D5 brane as the filling
fraction $\nu$ is increased to its critical value $\nu_c\sim 1.68f$ given in equation (\ref{criticalnu}) is seen when the D5 brane
is blown up to a D7 brane. In the above discussion, we have learned that in a certain region of the $(\nu,f)$ plane, the D7 brane
has lower energy than the D5 brane and is therefore the preferred solution.  Now we will report preliminary investigation of the 
boundaries of the region where it is more stable.  First of all, we have seen that, for moderate values of $f$ (in the range 1-10),  the D7 brane takes over from the D5 brane when $\nu$ becomes larger than $\frac{1}{2}$, at least in the region where the D5 brane would have a chiral symmetry breaking
solution with non-constant $\psi(r)$.   Thus, when $\nu<\frac{1}{2}$, at least in the regime that we can analyze, the standard D5 brane
phase diagram applies.  

The D5 embedding equation depends only on the ratio $\frac{\nu}{f}$.  
If we consider the conventional picture of the D5 brane  with a magnetic field and charge density,  at very small $\frac{\nu}{f}$, it is
in a chiral symmetry breaking state.  The function $\psi(r)$ has a non-constant profile which goes to $\frac{\pi}{2}$ as $r\to\infty$  (as
in equation (\ref{asymptoticpsi}) ) and goes to zero as $r\to 0$.  We are considering   only the case where 
$c_1=0$ and then, for a non-constant $\psi(r)$, $c_2$ must be nonzero.   If we increase the ratio  
$\frac{\nu}{f}$ the magnitude of the chiral condensate $c_2$ decreases and becomes zero when $\nu/f=\frac{2\sqrt{7}}{\pi}$.  At this critical
point, there is a phase transition with Berezinski-Kosterlitz-Thouless scaling \cite{Jensen:2010ga}. 
On the other side of the critical point, where $\nu/f>\frac{2\sqrt{7}}{\pi}$, the only solution of  the D5-brane theory is the constant
$\psi=\frac{\pi}{2}$.  

First, we observe that the D7 brane equation also has the same constant solution as the D5 brane equation $\psi=\frac{\pi}{2}$ with
the same energy and it becomes unstable at the same value of $\frac{\nu}{f}$.  This means that, whenever $\nu/f>\frac{2\sqrt{7}}{\pi}$ the constant $\psi=\frac{\pi}{2}$
is at least a meta-stable solution for the D7 brane.  

However, the D7 brane solutions with non-constant $\psi(r)$ have a different behavior than those for the D5 brane.  First of all, they depend
on both parameters $\nu$ and $f$, rather than just the ratio $\frac{\nu}{f}$.  Secondly, numerical computations show that they continue to
exist, even in the region $\frac{\nu}{f}>\frac{2\sqrt{7}}{\pi}\sim 1.68$ where the constant solution $\psi=\frac{\pi}{2}$ is a competitor.  
An example with $\nu=1$ and $f=0.1$, so that $\frac{\nu}{f}=10$,  is shown in figure \ref{fig14}.  
Not only does the solution exist there, but it has lower energy than the constant solution and it is therefore the energetically preferred configuration.  
Ungapped chiral symmetry breaking states are also energetically favorable in this region. 
Figure \ref{fig15} displays an example where $\frac{\nu}{f}=9$ and 
the D7 brane has a black hole embedding
and it is also stable.   In an attempt to find solutions which restore chiral symmetry, we have studied some examples with extreme values of 
$\frac{\nu}{f}$ where the D7 brane still has lower energy. Indications are that, for a very large range of parameters in the region $\nu>\frac{1}{2}$, the chiral symmetry breaking D7 brane
is the lowest energy solution. In the above, we
argued that the constant solution should be metastable in the whole region $\frac{\nu}{f}>\frac{2\sqrt{7}}{\pi}\sim 1.68$.  
The existence of non-constant solutions with lower energy there
implies that the constant and non-constant solutions are separated by an energy barrier.  This
in turn implies that, if their energies became equal or if the constant solution became energetically favorable in some region of the
parameter space, the chiral symmetry restoration phase transition would be first order.  At this point, we have not found numerical
evidence for such a phase transition,  however, this point certainly deserves additional study.

Recall that our Hall states for $\nu=1,2,\ldots,N_5$ have D7 branes with $f,f/2,...,f/N_5$ (assuming that the $N_5$ D5 branes are shared
symmetrically between the D7 branes) and therefore have $\frac{\nu}{f}=1,2,\ldots,N_5$.
There is still the open question as to how large this ratio can be before chiral symmetry is restored, or alternatively, how many of these quantum 
Hall states are stable.  As we have discussed above, we have found a numerical solution where $\frac{\nu}{f}=10$ (in figure \ref{fig14}) where the chiral
symmetry breaking Hall state is still the one which is energetically prefered.   We have also examined the extreme case of $\nu=1$ and $f=0.01$, that
is $\frac{\nu}{f}=100$ and it was also stable.

 \section{Conclusion}

We have formulated a mechanism by which integer quantum Hall states can appear in AdS/CFT holography which uses probe D5 branes
to construct strong coupling ${\cal N}=4$ Yang-Mills theory with fundamental representation matter fields occupying a supersymmetric 
2+1-dimensional defect.  The external magnetic field and finite
charge density that are needed in order to obtain a Hall state break supersymmetry (even in the free field limit, the fermions and bosons
have different Landau level spectra).  
We propose a number $N_5$ of D5 branes 
blown up to form a D7 brane as a candidate for the stable solution over a parameter range which we can easily
analyze when the number of D5 branes is large, $N_5\sim\sqrt{\lambda}$. 

 We find that for a range of filling fractions $\nu>0.5$ and for $f=\frac{2\pi}{\sqrt{\lambda}}N_5$ of order one-to-ten, the D7 brane has
lower energy than the D5 brane and it is therefore the preferred solution.  When $\nu<0.5$ the D5 brane becomes stable, at least
for the range of $f$ that we consider.  Some of the solutions with their energy densities are shown in figures \ref{fig5}-\ref{fig12}. 

When $\nu>0.5$ we find stable D7 brane solutions with non-constant $\psi(r)$ for a large range of values of the parameter $f$, 
including examples in the region where $\frac{\nu}{f}> \frac{2\sqrt{7}}{\pi}$.  This value of $\frac{\nu}{f}$ is a critical point of 
the D5 brane where there is a phase transition which restores chiral symmetry.   The situation is very different when $\nu>0.5$ where
the D7 brane is more stable.  In that case, the chiral symmetry breaking persists to larger values of $\frac{\nu}{f}$.  We have not
been able to determine whether there is a critical value of this parameter.  The competing chirally symmetric solution seems metastable
in this region, indicating existence of an energy barrier between the solutions.  Therefore one would expect a chiral symmetry restoring
phase transition to be of first order, if it exists at all.

We have used a comparison of energies to decide which D brane configurations are more stable.  However, we have not done a full
stability analysis.    
Particularly at finite density, instabilities to formation of non-translation invariant phases such as striped phases do occur for
the D7$^\prime$ system \cite{Bergman:2011rf} and such instabilities could also appear in the present system.  
We have left checking this important possibility for future work.

The mechanism by which we get higher filling fraction Hall states is by forming multiple D7 branes which share the total charge
density and which each have unit filling fraction $\nu=1$. In principle, states which have 
filling fractions $\nu=1,2,\ldots,N_5$ could exist.  How many of them exist is a dynamical question.  We have found numerical solutions
 where $\nu$ is as large as 10 and the Hall state still has the lowest energy.

We have used numerical computation to show that, for the first higher filling fraction state $\nu=2$, 
the two D7 branes with $\nu=1$ are stable in
that they have lower energies than the single D7 brane with $\nu=2$, which in turn has lower energy than the D5 branes with these
parameters.   The main reason for their stability 
is  that they are  Minkowski embeddings, so they gain energy by not having  spikes going to the Poincare horizon.  
Indications are that, in the multi D7 brane solution, at least for $\nu=2$, the state with D5 branes evenly distributed  between the two D7
branes is favored over asymmetric distributions.  The $\nu$ D7 branes are therefore coincident and have an SU($\nu$) symmetry. 
However, more work on mapping the energy landscape of multiple D7 branes is 
definitely needed before this is a firm conclusion.  
 
\section*{Acknowledgments}

C.F.K.\ 
was supported in part by FNU through grant number
272-08-0329.  G.W.S. is supported by NSERC of Canada and by the Villum
Foundation through their Velux Visiting Professor program. In
addition, G.W.S.  acknowledges the kind hospitality of the Niels Bohr
Institute and the University of Copenhagen.

\end{document}